\documentclass[graybox]{svmult} 

\usepackage{mathptmx}       
\usepackage{helvet}         
\usepackage{courier}        

\usepackage{amsmath}
\usepackage[utf8]{inputenc}     
\usepackage{enumerate}            
\usepackage[pdftex]{graphicx}             
\usepackage{tikz,pgf}
\usepackage{epstopdf}
\usepackage[
compatibility=false
]{caption}
\usepackage{subcaption}
\usepackage{wrapfig}
\usepackage{multicol}        
\usepackage[bottom]{footmisc} 
\usepackage[round]{natbib}

\makeindex
\begin{document}

\title*{
Criticality and self-organization in branching processes: application to natural hazards
}
\titlerunning{Criticality and self-organization in branching processes}
\author{\'Alvaro Corral and Francesc Font-Clos}
\authorrunning{\'Alvaro Corral and Francesc Font-Clos}
\institute{
\'Alvaro Corral \at
Centre de Recerca Matem\`atica, Edifici C, Campus Bellaterra, E-08193 Barcelona, Spain,
\email{acorral@crm.cat}
\and
Francesc Font-Clos \at
Centre de Recerca Matem\`atica, Edifici C, Campus Bellaterra, E-08193 Barcelona, Spain\\
Departament de Matem\`atiques, Universitat Aut\`onoma de Barcelona, E-08193 Barcelona, Spain
\email{fontclos@crm.cat}
}
\maketitle

\abstract{
The statistics of natural catastrophes contains very counter-intuitive results.
Using earthquakes as a working example, we show that the energy radiated
by such events follows a power-law or Pareto distribution.
This means, in theory, that the expected value of the energy does not exist
(is infinite), and in practice, that the mean of a finite set of data in not representative
of the full population. Also, the distribution presents scale invariance,
which implies that it is not possible to define a characteristic scale for the energy.
A simple model to account for this peculiar statistics is a branching process:
the activation or slip of a fault segment can trigger other segments to slip, with a certain probability,
and so on. Although not recognized initially by seismologists, this is a particular
case of the stochastic process studied by Galton and Watson one hundred years in advance,
in order to model the extinction of (prominent) families.
Using the formalism of probability generating functions we will be able
to derive, in an accessible way, the main properties of these models.
Remarkably, a power-law distribution of energies is only recovered in a very special case,
when the branching process is at the onset of attenuation and intensification, i.e.,
at criticality. In order to account for this fact, we introduce the self-organized critical models,
in which, by means of some feedback mechanism, the critical state becomes an attractor in
the evolution of such systems.
Analogies with statistical physics are drawn.
The bulk of the material presented here is self-contained,
as only elementary probability and mathematics
are needed to start to read.
}

\section{The Statistics of Natural Hazards}

\begin{quotation}
Only fools, charlatans and liars predict earthquakes\\
C. F. Richter\\
\end{quotation}


Men, and women, have always been threatened 
by the dangers of Earth: 
volcanic eruptions, tsunamis, earthquakes, hurricanes, floods, 
etc.
Sadly, still in the 21st century our societies have not been able
to get rid of such a sword of Damocles.
But
are natural catastrophes submitted to the caprices of the gods?
Or do these disasters contain some hidden patterns or regularities?
The first view has been dominant for many centuries
in the history of humankind, 
and it has been only in recent times that
a more rational perspective 
has started to consolidate.

\subsection{The Gutenberg-Richter law}

One of the first laws 
quantifying the occurrence of a natural hazard
was proposed for earthquakes
by the famous seismologists Beno Gutenberg and
Charles F. Richter in the 1940's,
taking advantage from the recent development of the first magnitude scale
by Richter himself.
The Gutenberg-Richter law is 
quite simple:
if one counts the number of earthquakes in any 
seismically active region of the world 
during a long enough period of time, one must find that
for each 100 earthquakes of magnitude $M$ greater or equal than 3
there are, approximately (on average), 10 earthquakes with $M\ge 4$,
one earthquake with $M\ge 5$,
and so on \citep{Gutenberg_Richter,Utsu_GR,Kanamori_rpp}.
So, the vast majority of events are the smallest ones, 
and, fortunately, only very few of them can become catastrophic,
maintaining a constant proportion between their number.

\begin{figure}
		\begin{subfigure}[b]{0.5\textwidth}
                \centering
                \includegraphics[height=8cm]{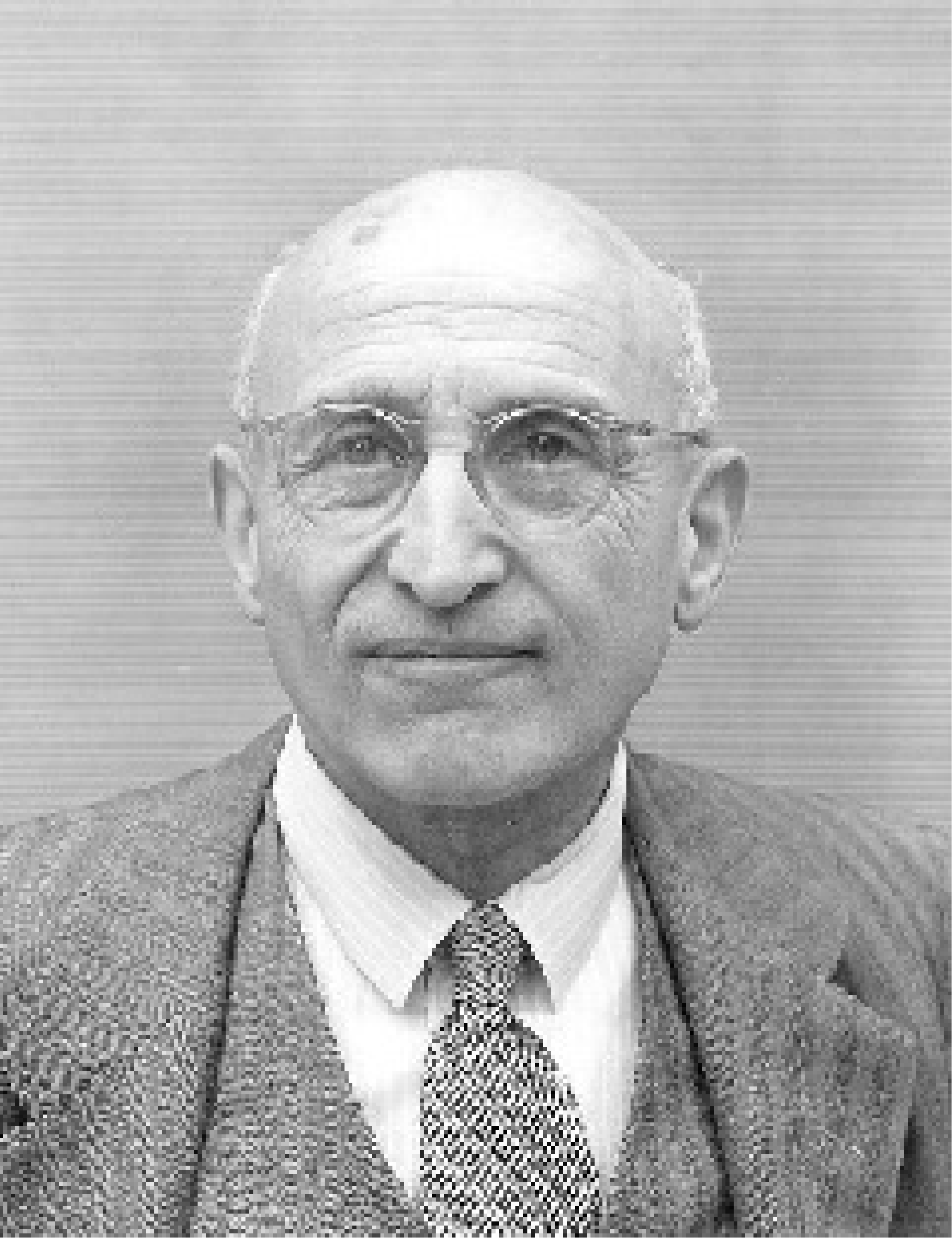}
                \caption{Beno Gutenberg}
        \end{subfigure}
        \begin{subfigure}[b]{0.5\textwidth}
                \centering
                \includegraphics[height=8cm]{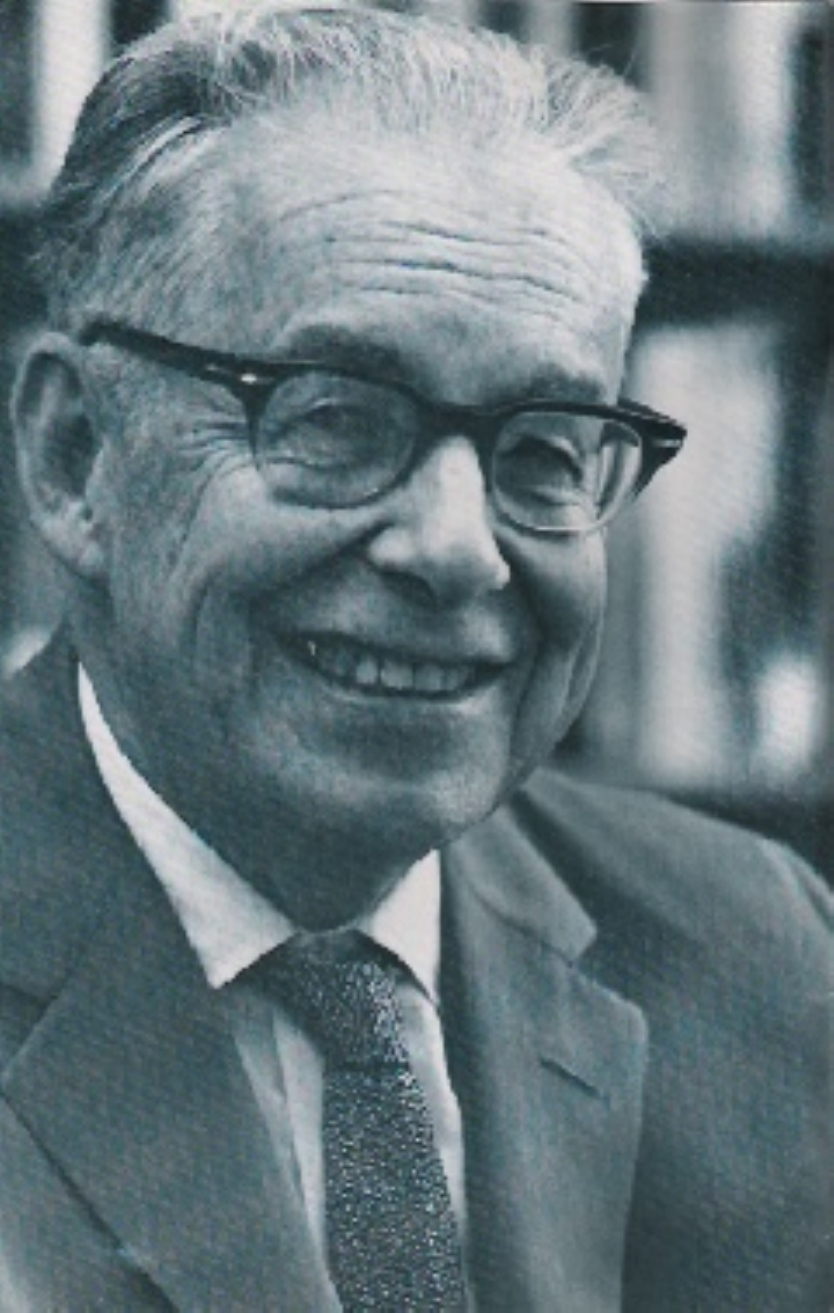}
                \caption{Charles Richter}
        \end{subfigure}
\caption{Seismologists Beno Gutenberg and Charles F. Richter
(photos from {\tt seismo.berkely.edu}).}
\label{single transition}
\end{figure}

It is not possible to measure all earthquakes on our planet,
but for some areas with very accurate seismic monitoring it has been found
that the Gutenberg-Richter law holds down to magnitude minus 4 \citep{Kwiatek};
this corresponds to small rock cracks of a few centimeters in length
(negative magnitudes are introduced to account for the fact
that there can be earthquakes smaller than those of zero magnitude).
And, more remarkably, 
for nanofracture experiments in the laboratory \citep{Astrom}, 
the law has been verified up to magnitude below -13.
The scarcity of the big events contained in the law
leaves as open the question about which is its upper limit of validity.

Despite not being recognized or mentioned
by Gutenberg and Ritchter in their original paper (1944),
any reader with a minimum knowledge of probability and statistics
will immediately realize that the Gutenberg-Richter law
implies an exponential distribution of the magnitudes
of earthquakes, i.e.,
\begin{equation}
D_M(M) \propto 10^{-b M}, 
\end{equation}
with $D_M(M)$ the probability density of $M$, 
the parameter $b$ taking a value close to 1, and
the symbol $\propto$ standing for proportionality
(with the constant of proportionality ensuring proper normalization).


But which is the meaning of the Gutenberg-Richter law, 
in addition to provide an easy-to-remember relationship
between the relative abundances of earthquakes?
The interpretation depends, of course, on the meaning
of magnitude, which we have avoided to define.
In fact, there is no a unique magnitude, but several of them, 
second, magnitudes do not have physical dimensions
(i.e., units), and third, 
``magnitudes reflect radiation only from subportions of the rupture, and they saturate
above certain size, rather than giving a physical characterization of the entire earthquake source''
\citep{Ben_Zion_review}. 
More in-depth understanding comes from the energy radiated
by an earthquake, which is believed to be 
an exponential function of its magnitude
\citep{Kanamori_rpp}, 
that is, 
\begin{equation}
\label{EofM}
E \propto 10^{3M/2},
\end{equation}
with a proportionality factor close to 60 kJ \citep{Utsu_GR};
so, an increase by 1 in the magnitude implies
an increase in energy by a factor $\sqrt{1000} \simeq 32$.
Thus, an earthquake of magnitude 9 radiates
as much energy as 1000 earthquakes of magnitude 7, or
as $10^6$ of magnitude 5.
 
One can reformulate then the Gutenberg-Richter law in 
terms of the energy.
Indeed, the probability of an event is ``independent''
of the variable we use to describe it, and so,
\begin{equation}
\label{DEtoDM}
D_E(E) = D_M(M)\frac{dM}{dE},
\end{equation}
with $D_E(E)$ the probability density of the energy.
Using equation \eqref{EofM}, we can express $M$ as a function of $E$,
\begin{equation}
M \propto \log E,
\end{equation}
and differentiate to obtain $dM/dE$,
\begin{equation}
\frac{dM}{dE} \propto \frac{1}{E},
\end{equation}
so that equation \eqref{DEtoDM} reads:
\begin{equation}
D_E(E) \propto 10^{-bM} \frac{1}{E}=
\left(10^{\frac{3M}{2}}\right)^{-\frac{2b}{3}}\frac{1}{E}=
E^	{-\frac{2b}{3}}\frac{1}{E}.
\end{equation}
Summarizing, 
this straightforward change of variables leads to 
\begin{equation}
\label{DEproptoE}
D_E(E) \propto \frac 1 {E^\alpha}, \,
\mbox{ with } \, 
\alpha = 1+ \frac{2b}{3},
\end{equation}
and this is just the so-called power-law distribution, 
or Pareto distribution \citep{Newman_05}, with exponent $\alpha$
around 1.67 when $b$ is close to 1.
Notice from equation \eqref{DEproptoE} that in order that $D_E(E)$ is a proper probability density function, 
it has to be defined above a minimum energy $E_{min}> 0$ ,
otherwise (if $E_{min}=0$), it cannot be normalized.
Although the true value of $E_{min}$
cannot be measured (it is too small),
this parameter is not important as it does not influence
any properties of earthquakes.

\begin{figure}
\centering
\hspace*{-15mm}
\includegraphics[width=0.7\textwidth]
{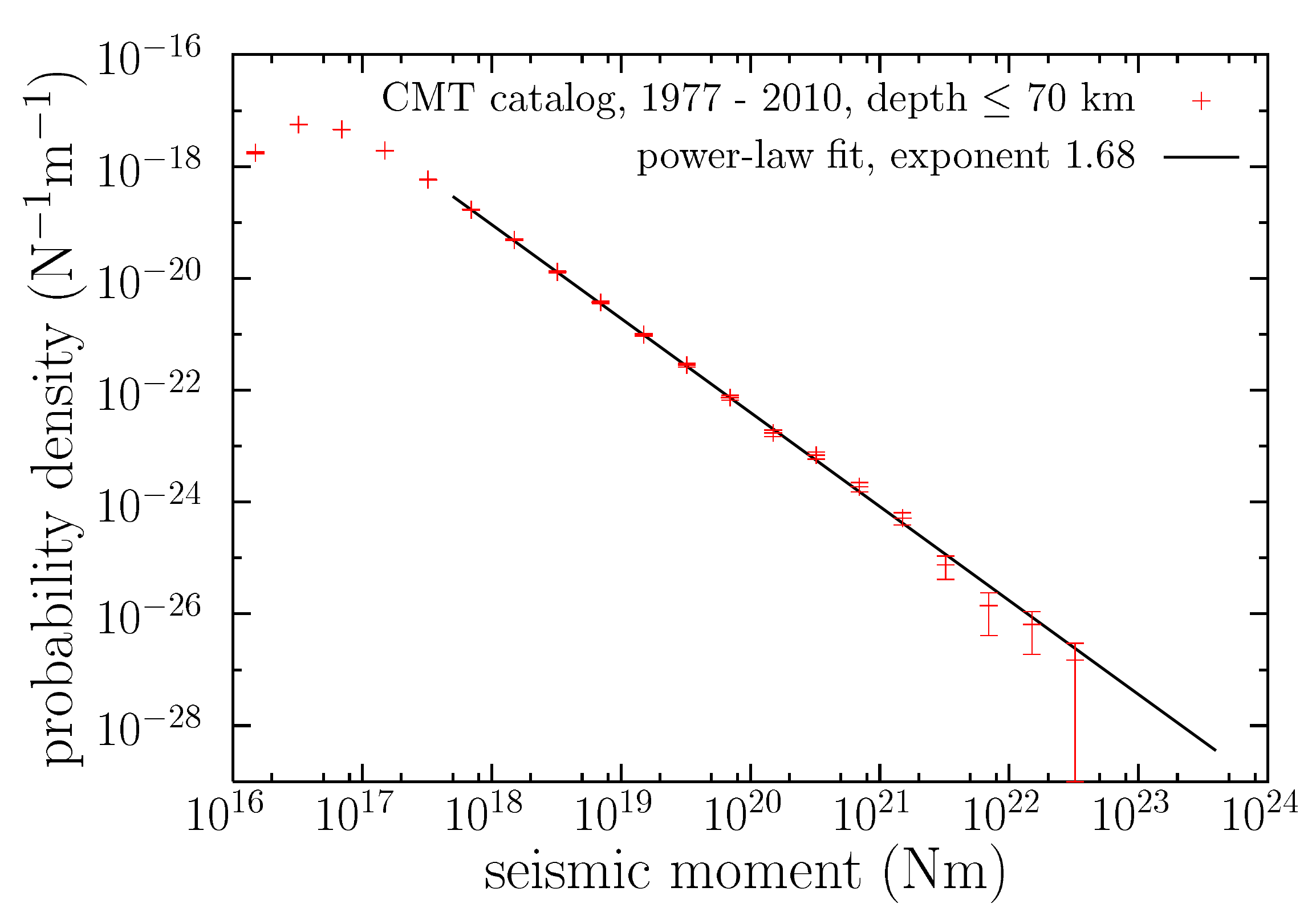}
\caption{Estimation of the probability density of seismic moment for worldwide shallow earthquakes
(in log-log scale), 
using the so-called CMT catalog \citep{Kagan_tectono10}. 
A power law fit results in an exponent $\alpha=1.68$. 
Radiated energy should give the same power law behavior.
Deviations at small values of the seismic moment are attributed to the incompleteness
of the catalog.}
 \label{seismic_moment} 
\end{figure}

Figure \ref{seismic_moment} 
displays the probability density of the seismic moment for worldwide shallow earthquakes
\citep{Kagan_tectono10};
this variable is assumed to be proportional to the energy,
but much easier to measure accurately \citep{Kanamori_rpp},
and so, it should also be power-law distributed, 
with the same exponent.
The straight line in the plot is the defining characteristic of a power law in
double logarithmic scale,
as $\log D_E(E)=C-\alpha \log E$.
A fit by maximum likelihood estimation
\citep{Clauset,Peters_Deluca} yields $\alpha \simeq 1.68$.

Two important properties of power-law distributions
are scale invariance (with some limitations due to 
the normalization condition) and divergence of the mean value
(if the exponent $\alpha$ is below or equal to 2). These are
explained in the Appendix.

To conclude this subsection, let us mention that the power-law distribution 
of sizes is not a unique characteristic of earthquakes.
It has been claimed that many other natural hazards are also power-law distributed,
although with different exponents
(and maybe with a lower or an upper cutoff):
tsunamis \citep{Burroughs}, landslides, rockfalls \citep{Malamud_hazards}, 
volcanic eruptions \citep{McClelland,Lahaie}, 
hurricanes \citep{Corral_hurricanes}, rainfall \citep{Peters_Deluca}, 
auroras \citep{Freeman_Watkins},
forest fires \citep{Malamud_fires_pnas}... 
As the reader will figure out, 
some of the facts that we will
explain having in mind earthquakes
can also be applied to some of these
natural hazards, but maybe not to all of them. It is an open question to 
distinguish between these different cases.
For an account of power-law distributions in other
areas beyond geoscience see the excellent review by \cite{Newman_05}.

%

\subsection{A first model for earthquake occurrence}

As far as we know, a first attempt to develop an earthquake
model in order to explain the Gutenberg-Richter law was undertaken
by Michio Otsuka in the early 1970's \citep{Otsuka71,Otsuka72,Kanamori_Mori}.
He used as a metaphor the popular Chinese game of go, 
although we will formulate the model in relation to the game of 
domino, probably more familiar to the potential readers.

Instead of playing domino, we are going to play a different
game with their pieces. The idea is to make the domino pieces
to topple, as in the well-known contests and attempts to break
a Guinness world record, but with two important differences.
First, the pieces are not put in a row, but, rather, they constitute
a kind of tree.
Second, when one piece topples, one does not know what will
happen next, i.e., if some other pieces will topple in turn 
(and how many will) or not.
So, we have a stochastic cascade process that supposedly mimics
the rupture that takes place in a seismic fault during an earthquake.
The tree of domino pieces constitutes the fault, and each piece is a small
fault patch, or element.
The earthquake is the chain reaction of toppling of pieces 
(i.e., failures of patches).

\begin{figure}
  \begin{center}
    \includegraphics[width=0.28\textwidth]{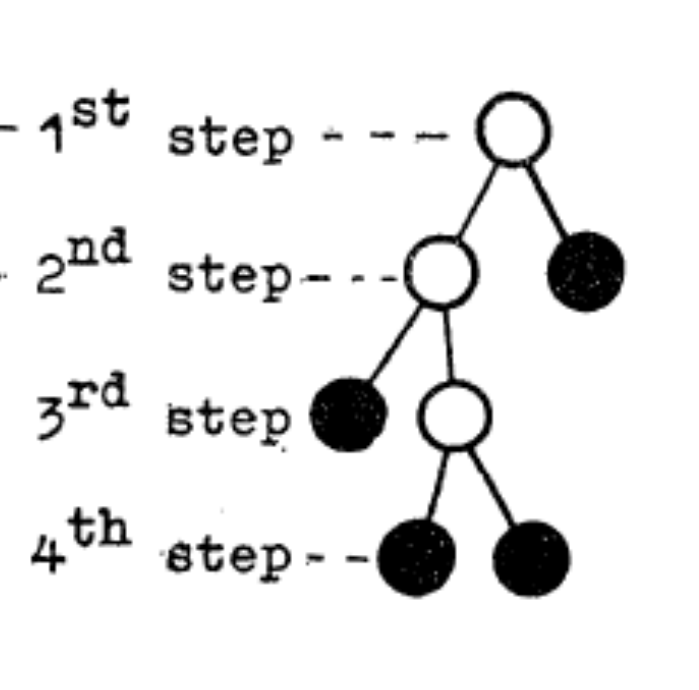}
  \end{center}
  \caption{Scheme of Otsuka's model for earthquake ruptures.
White circles correspond to the propagation of the rupture, 
whereas black ones indicate termination points  \citep{Otsuka72}.}
\label{figotsuka}
\end{figure}

Getting more concrete, Otsuka assumed that the tree representing the fault
had a fixed number of branches at each position, or node,
and that the toppling would propagate from each branch to the next
element with a fixed probability $p$,
independently of any other variable.
So, the number of propagating branches resulting from a single one 
would follow the binomial distribution \citep{Ross_firstcourse}.
For instance, in Fig. \ref{figotsuka}, the possible number of branches per element
is just 2.
If a fixed elementary energy is associated to the failure of each patch, 
one can obtain the energy released in this process from the number of
topplings, allowing the comparison with the Gutenberg-Richter law,
see nevertheless Sec. 4.1 of the review by \cite{Ben_Zion_review}.
So, the propagation of ruptures is considered a probability controlled phenomenon,
in such a way that when an earthquake starts, it is not possible to know
how big it will become.
Later, we will see that this statement is stronger 
than what it looks like here.
The usual domino effect, in which one toppling induces a new one
for sure and so on, would correspond to the controversial concept 
of a characteristic earthquake \citep{Stein_02,Ben_Zion_review,Kagan_rip}, 
an event that always propagates along the complete fault
or fault system and would release always the same amount of energy.

The novel and original model in geophysics explained in this
subsection, proposed by Otsuka in the 1970's, was
already known by a few mathematicians 100 years in advance.
It will take us the next pages to explain the distribution
of energy in this model.

    \section{Branching Processes}

\begin{quotation}
Besides gambling, many probabilists have been interested
in reproduction\\
G. Grimmett and D. Stirzaker
\end{quotation}

Let us move to the Victorian (19th century)
 England.
There, Sir Francis Galton, the polymath father of the statistical tools
of correlation and regression, and cousin of Charles Darwin,
was dedicated to many different affairs.
In addition to the height of sons in relation 
to the heights of their fathers, he
was concerned about the decay and even extinction of families
that were important in the past,
and about whether this decline was a consequence
of a diminution in fertility provoked by the rise in 
comfort.
If that were the case, 
population would be constantly fed
by the contribution of the lower classes
\citep{Galton_Watson}.
In order to better understand the problem, 
he devised a null model in which 
the number of sons of each men was 
random (the abundance of women was not considered to be a limitation).
Despite the apparent simplicity of the model, 
Galton was not able to solve it, and
made a public call for help.
The call was also fruitless, and then
Galton turned to the mathematician and reverend Henry William Watson.

\begin{figure}
		\begin{subfigure}[b]{0.5\textwidth}
                \centering
                \includegraphics[height=7cm]{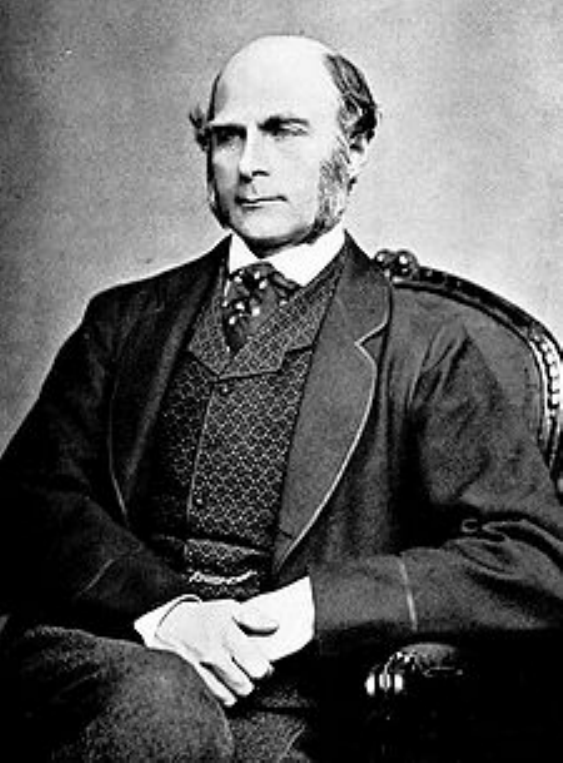}
                \caption{Sir Francis Galton}
                \label{fig:gull}
        \end{subfigure}
        \begin{subfigure}[b]{0.5\textwidth}
                \centering
                \includegraphics[height=7cm]{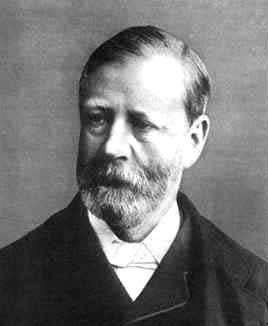}
                \caption{Rev. Henry William Watson}
                \label{fig:gull}
        \end{subfigure}
\caption{The fathers of the Galton-Watson process (photos from Wikipedia and {\tt www.wolframalpha.com}, respectively).}
\label{single transition}
\end{figure}

\subsection{Definition of the Galton-Watson process}

Let us consider ``elements'' that can generate other elements
and so on. 
These elements may represent British aristocratic men that have some male descendants,
(or, in a more fresh perspective, women from anywhere that give birth to her daughters,
or, perhaps more properly, bacteria that replicate),
neutrons that release more neutrons in a nuclear chain reaction, 
or fault patches that slip during an earthquake.
The Galton-Watson process assumes that 
each of these elements triggers a random number $K$ of offspring elements 
in such a way that each 
$K$
is independent from that of the other elements
and all $K$ are identically distributed,
with probabilities
$P(K=0)=p_0$, $P(K=1)=p_1$, 
\dots
$P(K=k)=p_k$, with $k=0,1, \dots \infty$ \citep{Harris_original}.
(Naturally, the normalization condition imposes
$\sum_{\forall k} p_k =1$.)

\begin{figure}[t]
\includegraphics[width=0.98\textwidth]{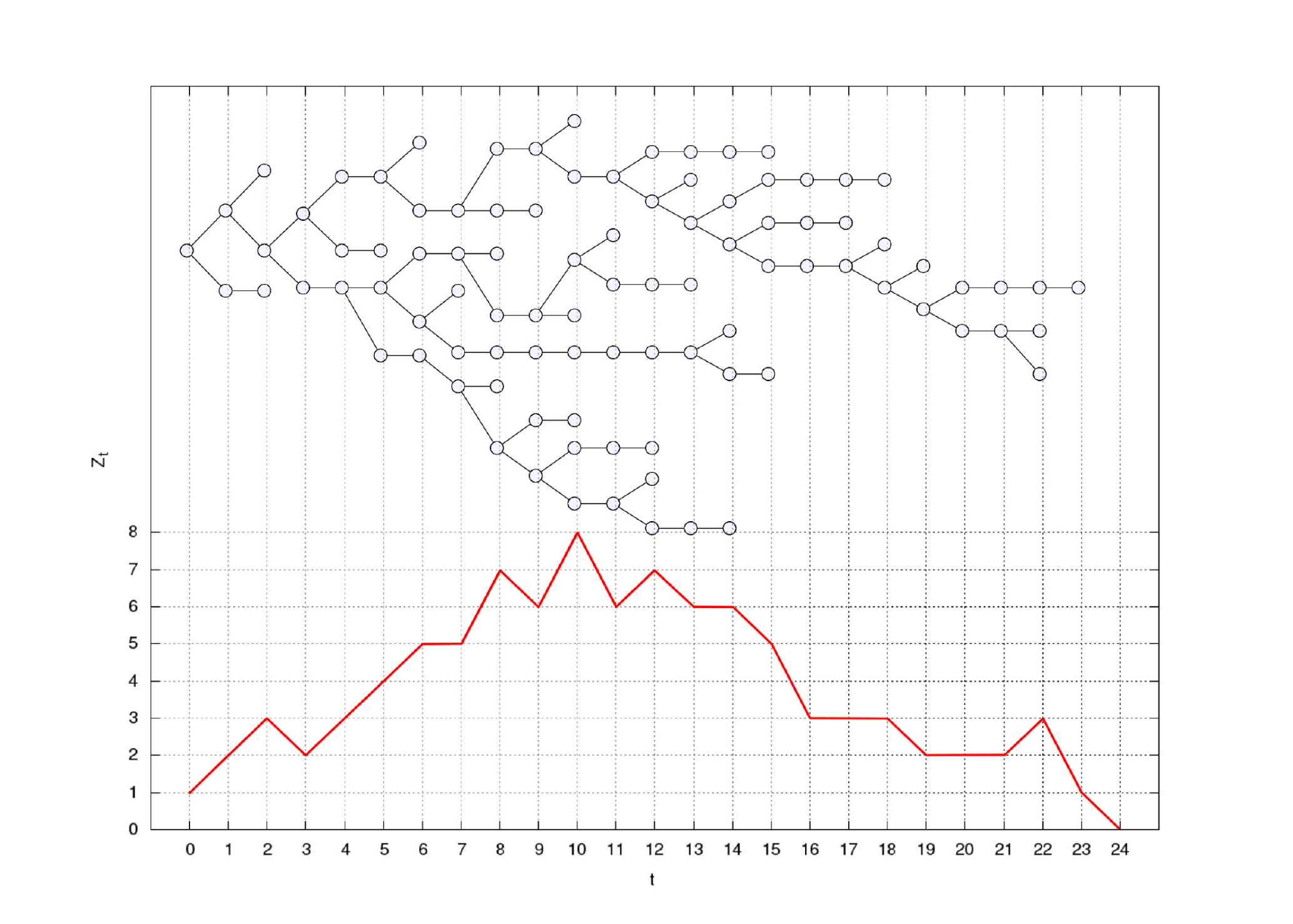}
\caption{A realization of the Galton-Watson process. At the top, 
the tree associated to the process is shown, starting from the left ($Z_0=1$).
At the bottom, the evolution of the number of elements originated in each generation $t$ are displayed.
The model for $P(K=k)$ is binomial with $n=2$ and $p=1/2$,
corresponding to the critical case (see main text).}
\label{example branching}
\end{figure}

The model starts with one single element, in what we call the
zeroth generation of the process, as shown in Fig. \ref{example branching}. The $K$ offsprings of this first element
constitute the first generation. Let $Z_0\equiv 1$ denote the number of elements
of the zeroth generation, $Z_1$ the number of elements of the first generation, etc.
Obviously, by construction, $P(Z_1=k)=p_k$.
The number of elements in the $t+1$ generation is obtained from the number 
of the previous generation $t$ as
\begin{equation}
\label{Zt+1}
Z_{t+1}=\sum_{i=1}^{Z_t} K_i,
\end{equation}
with $t \ge 0$,
where $K_i$ corresponds to the number of offsprings of 
each element in the $t$ generation.
Equation~\eqref{Zt+1} can be used to simulate the process in a straightforward way
and will be very important to its analytical treatment,
{ in order to calculate the probability distribution of $Z_t$, for any $t$}.
Some readers may recognize that the variables $Z_0, Z_1, \dots$
form a Markov chain, but this is not relevant for our purposes.
And of course, Otsuka's earthquake model is a particular case of the Galton-Watson process 
corresponding to a binomial distribution for $P(K=k)$.

\subsection{Generating functions}


An extremely convenient mathematical tool will be the probability generating function
\citep{Grimmett}.
For the random variable $K$ this is, by definition,
\begin{equation}
f_K(x) \equiv \sum_{k=0}^\infty p_k x^k = \langle x^K \rangle,
\end{equation}
where the brackets indicate expected value.
The normalization condition guarantees that
$f_K(x)$ is always defined at least in the $x-$interval $[-1,1]$,
although only the interval $[0,1]$ will be of interest 
for us.
Of course, the same definition applies to any other random variable;
in the concrete case of $K$ (which represents the number
of offsprings of any element) we may drop the subindex, 
i.e., $f_K(x)=f(x)$.

Very useful and straightforward properties will be,
\begin{enumerate}
\item $f_K(0)=P(K=0)$; 
\item $f_K(1)=1$ (by normalization);
\item $f_K'(1)=\sum_{\forall k}p_k k =\langle K \rangle \equiv m$;
\item 
$f_K'(x) \ge 0$ for $x\ge 0$ (non-decreasing function);

\item 
$f_K''(x) \ge 0$ for $x\ge 0$ (non-convex function, ``looking from above'');
\end{enumerate}
the primes denoting derivatives
(left-hand derivatives at $x=1$).
Note that although we illustrate these properties with the variable $K$,
they are valid 
for the generating function of any other (discrete) random variable.
So, the plot of a probability generating function between 0 and 1 is very constrained.
We anticipate that two main cases will exist, depending on whether
the expected value of $K$ is
$m<1$ or whether $m>1$.
This is natural, as the first case corresponds to a population
that on average decreases from one generation to the next
whereas in the second case the population grows, on average. 

Another property but not so straightforward is that the generating
function of a sum of $N$ independent identically distributed variables $K$
(with $N$ fixed)
is the $N$-th power of the generating function of $K$;
that is, if 
\begin{equation}
\Sigma=\sum_{i=1}^N K_i,
\end{equation} 
then
\begin{equation}
\label{fSigma1}
f_\Sigma(x) = f_K(x)^N.
\end{equation}
Indeed, 
\begin{equation}
f_\Sigma(x) = \langle x^\Sigma\rangle = \langle x^{\sum K_i}\rangle
= \langle x^{K_1} \cdot x^{K_2} \cdots x^{K_N} \rangle 
= \langle x^{K_1} \rangle \langle x^{K_2} \rangle \cdots \langle x^{K_N} \rangle   
= f_K(x)^N,
\end{equation}
where we can factorize the expected values due to statistical independence
among the $K_i$'s.

In general, if the random variables $K_i$ were not identically
distributed (but still independent),
the generating function of their sum would be the product of their 
generating functions. The demonstration is essentially the same 
as before, and one only needs to introduce new notation
for the different generating functions.

A following step is to consider 
that $N$ is also a random variable,
with generating function $f_N(x)$. Then,
\begin{equation}
\label{fSigma2}
f_\Sigma(x) = f_N(f_K(x)).
\end{equation}
Note that equation~\eqref{fSigma2} is just a generalization of equation~\eqref{fSigma1},
i.e., now we 
calculate the expected value of
the powers of $f_K(x)$ depending on the values 
that $N$ make take. In any case, it is easy to demonstrate:
{ denoting with $\langle \cdot \rangle_{K_i}$ the average over the $K_i$'s} 
and with $\langle \cdot \rangle_N$ the average over $N$, we have

\begin{equation}
f_{\Sigma}(x) =\langle x^{\Sigma} \rangle = 
\left\langle \langle x^{\Sigma}\rangle_{K_i} \right\rangle_{N} = \Bigl\langle f_{K}(x)
^{N} \Bigr\rangle_{N} = f_{N}(f_{K}(x)),
\end{equation} 
where the last equality is just the definition of the probability generating function 
of the random variable $N$, {evaluated at $f_K(x)$}. 
We stress that this is only valid for independent random variables.

\subsection{Distribution of number of elements per generation} 

Going back to the Galton-Watson branching process, 
where we know that $Z_{t+1}=\sum_{i=1}^{Z_t}K_i$,
we can identify $Z_{t+1}$ as $\Sigma$ and $Z_t$ as $N$; then equation \eqref{fSigma2} reads,
\begin{equation}
f_{Z_{t+1}}(x)=f_{Z_{t}}(f_{K}(x))=f_{Z_{t}}(f(x))
\end{equation}
(dropping the subindex $K$).
As $f_{Z_1}(x)=f(x)$,
it is straightforward to see by induction that the generating function of $Z_{t}$, 
is given by
\begin{equation}
\label{compof}
f_{Z_{t}}(x) = f(f(...f(x)))=f^{t}(x),
\end{equation}
where the superindex $t$ denotes composition $t$ times.
This is valid for $t=1,2, \dots$; for $t=0$ we have, obviously, that
$f_{Z_{0}}(x)=x$ (because $Z_0=1$ with probability 1). 
In words, the generating function of the number of elements for each
generation is obtained by the successive compositions
of $f(x)$.
This non-trivial result was 
first proved by Watson in 1874 \citep{Harris_original}.

\subsection{Expected number of elements per generation }

Here we present an illuminating result, 
which will be useful at some point in the chapter.
Although, in general, the successive compositions of the generation function 
leads to very complicated mathematical expressions,
the moments of $Z_t$ can be computed in a simple way \citep{Harris_original}.
Using what we have learnt about generating functions together wtih equation~\eqref{compof}, 
the expected value of $Z_t$ is
\begin{equation}
\langle Z_t \rangle =  \left. \frac {d}{dx} f^t(x)\right |_{x=1}.  
\end{equation}
Let us then write 
\begin{equation}
\frac {d}{dx} f^t(x)=\frac {d}{dx} f(f^{t-1}(x))=
f'(f^{t-1}(x)) \frac {d}{dx} f^{t-1}(x),
\end{equation}
therefore, by induction,
\begin{equation}
\frac {d}{dx} f^t(x)=f'(f^{t-1}(x)) 
f'(f^{t-2}(x)) \cdots f'(f^{2}(x)) f'(f(x))f'(x).
\end{equation}
Taking $x=1$ and using that all the generating functions
have to be 1 at that point,
\begin{equation}
\langle Z_t \rangle = f'(1)^t = m^t.
\end{equation}
So, when $m<1$ the mean number of elements per generation decreases exponentially,
whereas when $m>1$ this number increases, 
constituting a stochastic realization of Malthusian growth.
For this reason $m$ is sometimes called the branching ratio.
When $m=1$ the average size of the population is constant,
but we will later see that this does not mean that the population
reaches a stable state.
Higher-order moments can be computed in a similar way, 
but they are not so useful as the mean.

Another related issue is the one of the expected value
of the number of elements per generation conditioned
to the value of the previous generation, 
i.e., ${\langle Z_{t+1} | Z_t=z_t\rangle}$.
As when $Z_t$ is fixed, $Z_{t+1}=\sum_{i=1}^{z_t} K_i$, 
then, taking the expected value,
\begin{equation}
\langle Z_{t+1} |  Z_t=z_t\rangle = \sum_{i=1}^{z_t} \langle K_i \rangle = z_t m. 
\end{equation}
This result can be used to relate branching processes
with martingales \citep{Grimmett}, 
but this does not have to bother us.



\subsection{The probability of extinction}

Extinction of the process is achieved when $Z_t=0$,
for the first ``time''
(i.e., for the generation that yields $Z_t=0$ for the first $t$).
Then, all the subsequent $Z$'s are also zero, 
and extinction can be considered an ``absorbing state'',
in this sense.
We now see that the probability of extinction in the Galton-Watson
process is equal to one (extinction for sure) for $m\le 1$
and is smaller than one for $m>1$.

This result, which may be referred to as the Galton-Watson-Haldane-Steffensen 
(criticality) theorem, was first proved by J. F. Steffensen, in the 1930's
(being unaware of the work by Galton and Watson,
and later progress by Haldane).
As \cite{Kendall66} pointed out, after then,
the same theorem ``was to be re-discovered over and over again, especially
during the (Second World) War period, and no doubt we have not yet seen its last
re-discovery''.
Ironically, Kendall did not know that Ir\'en\'ee-Jules Bienaym\'e 
knew the theorem, in its correct formulation, 30 years in advance Galton and Watson
and 85 years before Steffensen
\citep{Kendall75}!

Indeed, extinction may happen at the first generation, $Z_1=0$,
or at the second, $Z_2=0$, etc.
All these extinction events are included in $Z_t=0$, with $t\rightarrow \infty$;
therefore, the probability of extinction $P_{ext}$ is given by
\begin{equation}
P_{ext}=
\lim_{t\rightarrow \infty}P(Z_1=0 \mbox{ or } Z_2=0 \mbox{ or } \dots \mbox{ or } Z_t=0)=
\lim_{t\rightarrow \infty} P(Z_t=0)=\lim_{t\rightarrow \infty} f^t(0),
\end{equation}
i.e., by the infinite iteration of the point $x=0$ through the 
generating function $f(x)$
(using the key property that the probability of a zero value
is the value of the generating function at zero,
and equation~\eqref{compof} again).

\begin{figure}
\includegraphics[width=\textwidth]{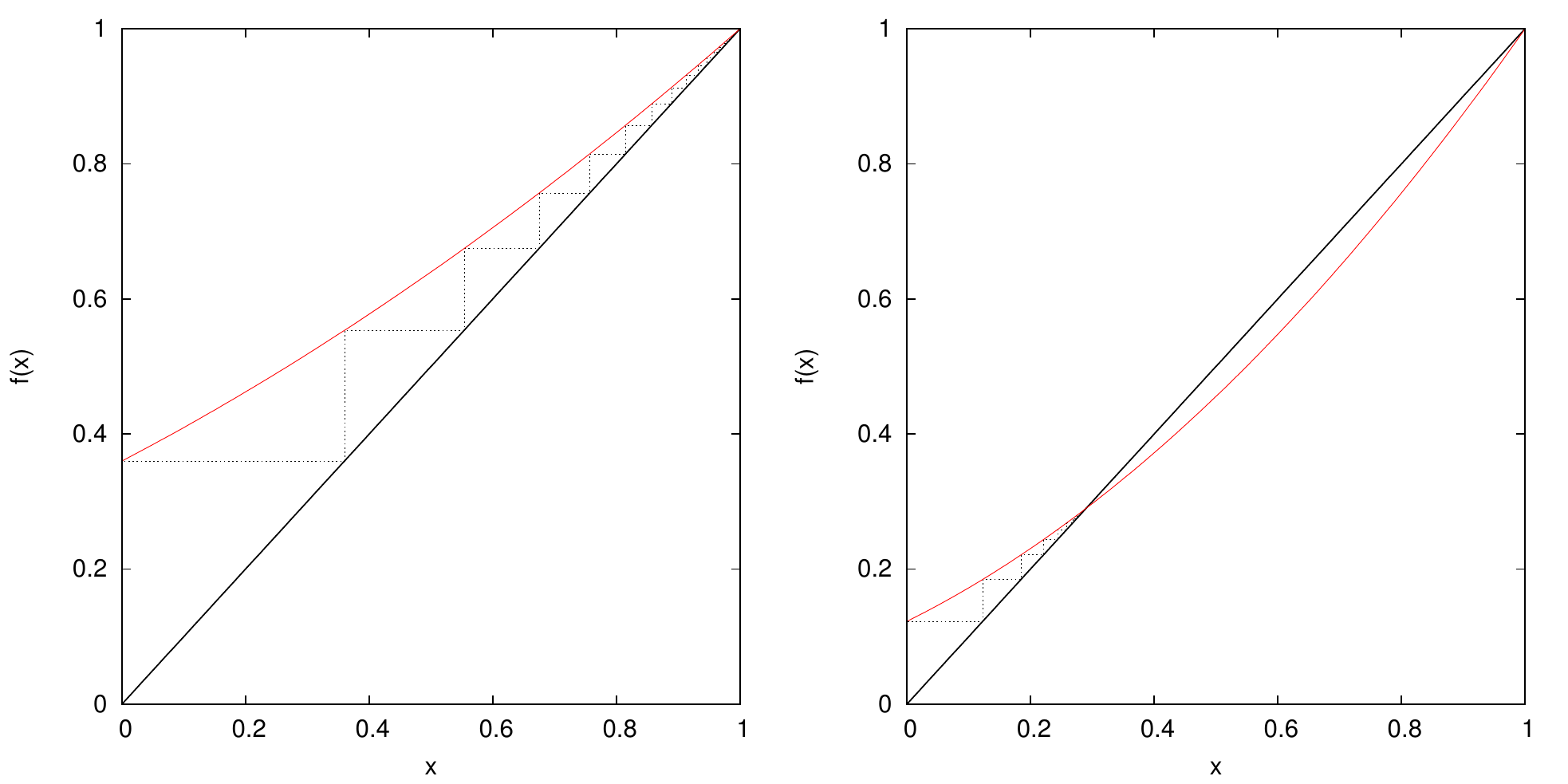}
\caption{Probability generating function $f(x)$ of the number of offsprings per element
and iteration of the point $x=0$ through successive compositions of $f$.
The fixed points correspond to the crossings of the diagonal;
those closer to zero are also the attractors for the iteration.
Left corresponds to a subcritical case and right to a supercritical case.
The model is the binomial one, with $n=2$.
}
\label{iterate}
\end{figure}

We now calculate the iteration $f^t(0)$.
In the interval $[0,1]$ 
the function $f(x)$ is 
non-decreasing
and non-convex,
taking values from $p_0$ to 1. 
If the slope of $f(x)$ at $x=1$, given by 
$m=\langle K \rangle =f'(1)$, is smaller than or equal to 1,
then $f(x)$ only crosses (or reaches) the diagonal at $x=1$
({otherwise, $f(x)$ would need to be convex somewhere}), 
and the iteration of the point $x=0$ ends at the point $x=1$ (which is the attractor,
see Fig. \ref{iterate}).
Therefore, 
\begin{equation}
P_{ext}=\lim_{t\rightarrow \infty} f^t(0)=1,
\end{equation}
i.e., extinction is unavoidable if $m\le 1$.
{
There is a trivial exception, though, associated to $p_1=1$ (and zero for the
rest); this is an extremely boring situation indeed.
In this case, $f(x)=x$, and therefore $\lim f^t(0)=0$,
which means, obviously, that the probability of extinction is zero.
}

If the slope of $f(x)$ at $x=1$ is $m>1$
(which only can happen for a non-linear generating function, $p_0+p_1< 1$), 
then $f(x)$ has to cross the diagonal 
at a point $x^*$ smaller than one, which is the attractive solution to which 
the iteration tends, see Fig. \ref{iterate} again. In mathematical language, 
\begin{equation}
P_{ext}=\lim_{t\rightarrow \infty} f^t(0)=x^*,
\end{equation}
where
\begin{equation}
\label{fixedpoint}
x^*=f(x^*) \, \mbox { with } x^* <  1.
\end{equation}
The 
demonstration is elaborated in the Appendix.

Summarizing,
\begin{equation}
\label{Pext}
P_{ext}= \left \{
\begin{array}{lr}
1 & \mbox{ if } m \le 1 \\
x^* & \mbox{ if } m > 1 \\
\end{array}
\right.
\end{equation}
with $x^* <  1$,
except in the trivial case $p_1=1$, 
which has $m=1$ but 
yields $P_{ext}=0$.

Equation~\eqref{Pext} clearly shows that, in general, the point $m=1$ separates
two distinct behaviors: extinction for sure for $m\le 1$
and the possibility of non-extinction (non-sure extinction)
for $m>1$. Therefore, $m=1$ constitutes a critical case
separating these behaviors,
called therefore subcritical ($m<1$)
and supercritical ($m>1$).
It is instructive to point out that, as $x=1$ is always a solution of
$f(x)=x$, Watson concluded, incorrectly, that the population always
gets extinct, no matter the value of $m$ \citep{Kendall66}.

\subsection{The probability of extinction for the binomial distribution}

For the sake of illustration
we will consider a simple concrete example, a binomial distribution \citep{Ross_firstcourse,Grimmett},
\begin{equation}
p_k =P(K=k)=\left(
\begin{array}{c}
n \\ k \\
\end{array}
\right)
p^k (1-p)^{n-k},
\, \mbox{ for } \, k=0, \dots n.
\end{equation}
This assumes that each element has only a
fixed number of trials $n$
to generate other elements, 
and any of these $n$ trials 
has a constant probability $p$ of being successful.
The generating function turns out to be, using the
binomial theorem
\begin{equation}
f(x)=\sum_{k=0}^\infty 
\left(
\begin{array}{c}
n \\ k \\
\end{array}
\right)
 (1-p)^{n-k} p^k x^k=
(1-p + px)^n.
\end{equation}

Let us consider 
the simple case with $n=2$,
and define $q=1-p$.
As we know,
the probability of extinction will come from 
the smallest solution in $[0,1]$ of
\begin{equation}
x=(q+px)^2.
\end{equation}
So,
\begin{equation}
x=\frac{1-2pq \pm \sqrt{(1-2pq)^2-4p^2q^2}}{2p^2},
\end{equation}
but the square root can be written
as $\sqrt{1-4p(1-p)}=\sqrt{(1-2p)^2}=(1-2p)$,
and then,
\begin{equation}
x=\frac{1-2p +2p^2 \pm (1-2p)}{2p^2}=
\left\{
\begin{array}{c}
\left(\frac q p\right)^2\\
1\\
\end{array}
\right.
\end{equation}
Therefore, the smallest root depends on whether $p$
is below or above $1/2$
\begin{equation}
\label{Pext2}
P_{ext}=
\left\{
\begin{array}{cc}
1 & \mbox{ for } p \le \frac 1 2\\
\left(\frac q p\right)^2 & \mbox{ for } p \ge \frac 1 2 \\
\end{array}
\right.
\end{equation}
As for the binomial distribution $m=np=2p$ \citep{Ross_firstcourse},
the critical case $m=1$ corresponds obviously to $p=1/2$,
in agreement with the behavior of $P_{ext}$.

\subsection{No stability of the population}

\begin{figure}
\includegraphics[width=\textwidth]{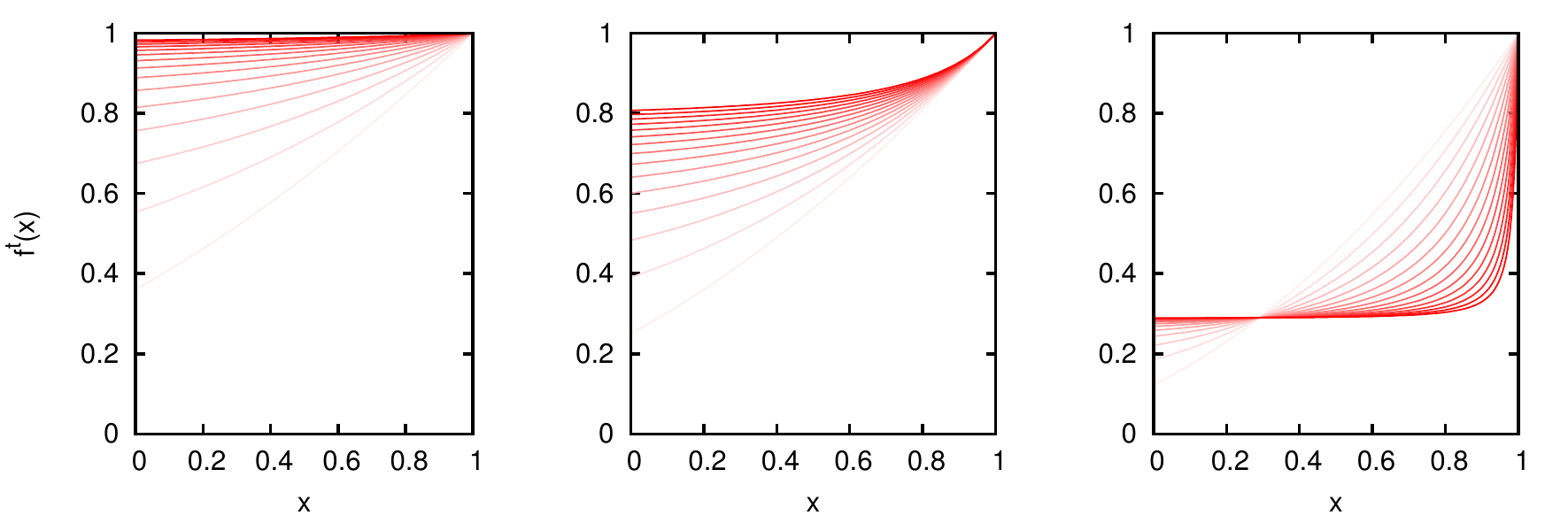}
\caption{Successive compositions of $f(x)$, for all $x$, 
yielding the probability generating functions of $Z_t$,
starting at $t=1$ (lighter red) up to $t=15$ (darker red). Larger $t$ leads to flatter functions,
approaching the fixed point.
From left to right, subcritical, critical, and supercritical cases, 
using a binomial model with $n=2$.
}
\label{iterate2}
\end{figure}

Although this subsection contains an interesting result
to better understand  the behavior of the Galton-Watson process, 
it can be skipped as it is not connected to the rest of 
the chapter.
In fact, the iteration of the point $x=0$
shows what happens to the whole generating function
of $Z_t$ when $t\rightarrow \infty$.
Indeed,
in the same way as in subsection 2.5,
\begin{equation}
\lim_{t\rightarrow \infty} f_{Z_t}(x)=
 \lim_{t\rightarrow \infty} f^t(x)=1 \, \mbox{ if } m\le 1,
\end{equation}
whereas 
\begin{equation}\lim_{t\rightarrow \infty} f_{Z_t}(x)=
 \lim_{t\rightarrow \infty} f^t(x)=x^* < 1 \, \mbox{ if } m > 1,
\end{equation}
except for $x=1$, 
which always fulfills $ \lim_{t\rightarrow \infty} f^t(x)=1$,
see Fig. \ref{iterate2}).

Note that a flat generating function corresponds
to probabilities equal to zero, except for the zero value,
i.e., \begin{equation}
\lim_{t\rightarrow \infty} P(Z_t=k)=0, \,
\mbox{except for } k=0.
\end{equation}
In this way, for $m\le 1$ we have that
$\lim_{t\rightarrow \infty} P(Z_t=0)=1$,
and the population gets extinct;
but for $m> 1$ we have found
$\lim_{t\rightarrow \infty} P(Z_t=0)=x^* < 1$;
having any other finite value of $K$ a zero probability, 
this means that $Z_t$ goes to infinite, when $t \rightarrow \infty$,
with probability $1-x^*$;
that is, $Z_t$ cannot remain positive and bounded.
The only stable state is extinction.
Obviously, in this limit the Galton-Watson process is unrealistic,
as other external factors should prevent that the population goes to
infinity.
But we do not need to bother about that, 
if we understand the limitations of the model.

\subsection{Non-equilibrium phase transition}

Let us analyze in more detail what happens around the ``transition 
point'' ${m=1}$.
As we just have seen, recall equation~\eqref{fixedpoint}, the extinction probability is given by the
solution of ${P_{ext}=f(P_{ext})}$. 
When $m\le 1$ the only solution in $[0,1]$ is $P_{ext}=1$
(except in the trivial case $p_1=1$). 
When $m >1$ we have to take the smallest solution of $P_{ext}=f(P_{ext})$
in $[0,1]$.
In terms of the non-extinction probability, $\rho=1-P_{ext}$, 
we need to look for the largest $\rho$ that is solution
of
\begin{equation}
f(1-\rho)= \sum_{k=0}^\infty p_k (1-\rho)^k = 1- \rho,
\end{equation}
in the range $[0,1]$.
We explore the case of
$P_{ext}$ close to 1, for which $\rho$ is close to zero, 
and, using the binomial theorem, we can expand 
$(1-\rho)^k = 1-k\rho + k(k-1)\rho^2/2 + \cdots$,
which yields
\begin{align}
\sum_{k=0}^\infty p_k - \sum_{k=0}^\infty k p_k \rho 
+ \frac 1 2 \sum_{k=0}^\infty  k(k-1)p_k \rho^2  + \cdots  &=
\\
\notag =1 - m\rho + \frac 1 2 \mu \rho^2 + \cdots &= 1-\rho,
\end{align}
where we have introduced the mean $m$ and the second 
factorial moment $\mu=\langle K(K-1) \rangle$ (which we assume exists).
Therefore, up to second order in $\rho$ we need to solve
\begin{equation}
\label{secondorder}
\left(\frac 12 \mu \rho + 1-m \right) \rho \simeq 0.
\end{equation}
It is immediate that one solution of equation~\eqref{secondorder} is $\rho=0$,
and one can realize that this solution is exact up to any order in $\rho$.
The other solution is $\rho \simeq 2 (m-1)/\mu$, but we must pay attention
to the value of $\mu$, which can be written as $\mu=\sigma^2+m(m-1)$, 
with $\sigma^2=\langle (K-m)^2 \rangle=\langle K^2 \rangle -m^2$,
i.e., the variance. 
Existence of $m$ and $\sigma^2$ guarantees the existence of $\mu$, then.
Assuming $\sigma^2 \ne  0$,
\begin{equation}
 \frac {2(m-1)} \mu = \frac {2(m-1)} {\sigma^2 [ 1 + m (m-1)/\sigma^2]}
= \frac {2(m-1)} {\sigma^2} \left[ 1- \frac{m(m-1)}{\sigma^2} + \dots\right]
\end{equation}
(using the formula for the geometric series),
therefore, $\rho$ around zero means $m$ around one, and we can write
the second solution as
\begin{equation}
\rho \simeq \frac {2(m-1)} {\sigma^2}
\end{equation}
which is only in the range of interest for $m > 1$.

%

In conclusion, we have
\begin{equation}
\label{readoff}
\begin{array}{ll}
\rho = 0 & \mbox{ if } \, m \le 1 \\
\rho \simeq {2 (m-1)/\sigma^2}  & \mbox{ if } \, m > 1, \\
\end{array}
\end{equation}
valid in the limit of small $\rho$. For $m>1$ this limit
is equivalent to $ m \rightarrow 1$.
The separate case $\sigma^2=0$ is only achieved in the trivial situation where
$p_1=1$ (otherwise, the mean cannot approach one).

In this way, we obtain a behavior that is the one corresponding to a 
continuous phase transition in thermodynamic 
equilibrium. Identifying $m$ with a control parameter
(as temperature, or more properly, the inverse of temperature) 
and $\rho$ with an order parameter
(as magnetization in a magnetic system) these transitions show 
an abrupt but continuous change of $\rho$ as a function of $m$
at the transition point $m_c$, with 
\begin{equation}
\begin{array}{ll}
\rho=0  & \mbox{ below } \, m_c \\
\rho \propto (m-m_c)^\beta & \mbox{ above but close to} \, m_c \\
\end{array}
\end{equation}
For magnetic systems, $m_c$ corresponds to the so-called Curie temperature.
For the Galton-Watson branching process we can extract from equation~\eqref{readoff} that
\begin{equation}
m_c=1 \, \mbox{ and } \beta=1,
\end{equation}
where we assume that the variance of $K$ does not go to zero 
at the transition point.

We can compare the previous general result,
$\rho \simeq 2(m-1)/\sigma^2$, for $m$ above but close to 1,
with the result we found for the binomial distribution
with $n=2$ (see equation~\eqref{Pext2}), for which 
\begin{equation}
\rho=1-\left(\frac{1-p}{p}\right)^2 = \frac{2p-1}{p^2}
\end{equation} 
when $p\ge 1/2$.
Using that in this case $m=np$ and $\sigma^2=npq$ (see \cite{Ross_firstcourse}),
\begin{equation}
\frac{2(m-1)}{\sigma^2} = \frac{2p-1}{pq} \simeq \frac{2p-1}{p^2},
\end{equation}
because $q=1-p \simeq p$ for $p\simeq 1/2$.
So, equations~\eqref{Pext2} and \eqref{readoff} agree close to the transition point.
Figure \ref{single transition} shows also how they disagree as $m$ increases.

\begin{figure}
		\begin{subfigure}[b]{0.5\textwidth}
                \centering
                \includegraphics[width=\textwidth]{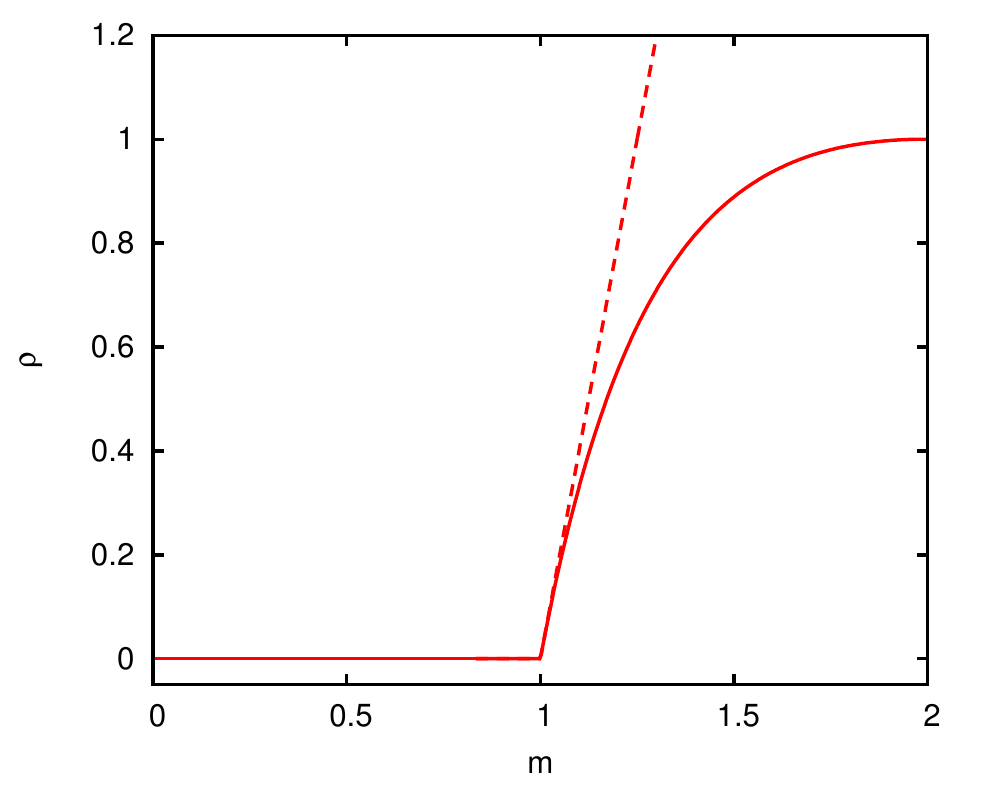}
                \label{fig:gull}
        \end{subfigure}
        \begin{subfigure}[b]{0.5\textwidth}
                \centering
                \includegraphics[width=\textwidth]{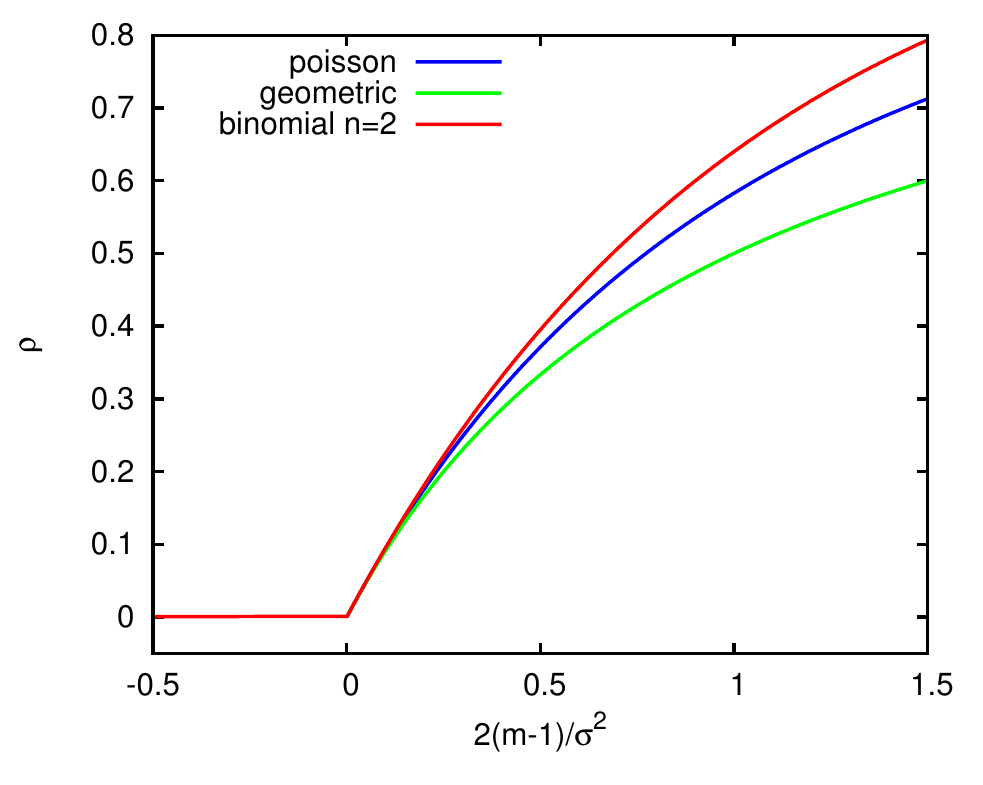}
                \label{fig:gull}
        \end{subfigure}
             \caption{Left: non-extinction probability $\rho$ as a function of the mean number of offsprings per element, $m$.
Dashed line corresponds to the approximation explained in the text (eq.~\eqref{readoff}).
The abrupt change in $\rho$ is the hallmark of a continuous phase transition.
The model is binomial with $n=2$.
Right: the same but as a function of the rescaled distance to the critical point, 
$2(m-1)/\sigma^2$, where $\sigma^2$ refers to the variance at $m=1$.
The Poisson and the geometric distributions are also studied. 
}
\label{single transition}
\end{figure}



Finally, for completeness, we can play with the pathological case given by
$\sigma^2=0$.
Let us consider first the following model, 
$p_0=1-\lambda_1$, $p_1=\lambda_1$ (and zero otherwise),
with $\lambda_1 < 1$.
Then, $m=\lambda_1$, and we know that $\rho=0$.
Next, let us consider 
$p_1=1-\lambda_2$, $p_2=\lambda_2$ (and zero otherwise),
giving $m=1+\lambda_2$.
In this case, $\rho=1$ always,
yielding a discontinuous, or first order phase transition.

\subsection{Distribution of the total size of the population: binomial distribution and rooted trees}

Our main interest will now be to calculate the total size $S$ of the population,
summing across all generations, i.e.,
\begin{equation}
S=\sum_{t=0}^\infty Z_t,
\end{equation}
this corresponds to the total number of individuals that have ever been born, 
the total number of neutrons participating in a nuclear chain reaction,
or the energy released during an event in an earthquake model.

Let us go back to the concrete binomial case,
\begin{equation}
p_k =P(K=k)=\left(
\begin{array}{c}
n \\ k \\
\end{array}
\right)
p^k (1-p)^{n-k},
\, \mbox{ for } \, k=0, \dots n.
\end{equation}
The size distribution can be calculated using elementary probability and
combinatorics.
One needs to take advantage of the representation of a branching process
as a tree (which is a connected graph with no loops). 
Each element is associated to a node, and branches linking nodes
indicate an offspring relationship between two nodes.
Naturally, all nodes have just one incoming branch, 
except the one corresponding to
the zero generation (which in this context is called the root
of the tree). 
So, the number of branches is the number of
nodes minus 1.
As the size $s$ of a tree is the number of nodes it contains, 
the number of branches is $s-1$,
and the number of missing branches (non-successful reproductive trials)
is $ns-(s-1)$ (because the number of possible branches arising from 
$s$ nodes is $ns$) \citep{Christensen_Moloney}.
Therefore, a particular tree of size $s$ comes with a probability
$p^{s-1}(1-p)^{(n-1)s +1}$, and the probability $P(S=s)$ of having an undefined
tree of size $s$ is obtained by summing for all possible trees of size $s$.
In the case $n=2$ the number of 
trees with $s$ nodes is given
by the Catalan number 
\begin{equation}
C_s = \frac 1 {s+1} 
\left(
\begin{array}{c}
2s \\ s \\
\end{array}
\right),
\end{equation} 
see the Appendix for its calculation. Then,
\begin{equation}
\label{catalan1}
P(S=s) = \frac 1 {s+1} 
\left(
\begin{array}{c}
2s \\ s \\
\end{array}
\right)
p^{s-1}(1-p)^{s +1}
\, \mbox{ with } \, 
s=1,2,\dots
\end{equation}
It can be checked, using the generating function
of the Catalan numbers, that 
this expression is normalized for $p\le 1/2$ but not for $p> 1/2$,
in fact,
\begin{equation}
\sum_{s=1}^\infty P(S=s)= P_{ext},
\end{equation}
see the Appendix again.

\begin{figure}[h]
	\includegraphics[width=\textwidth]{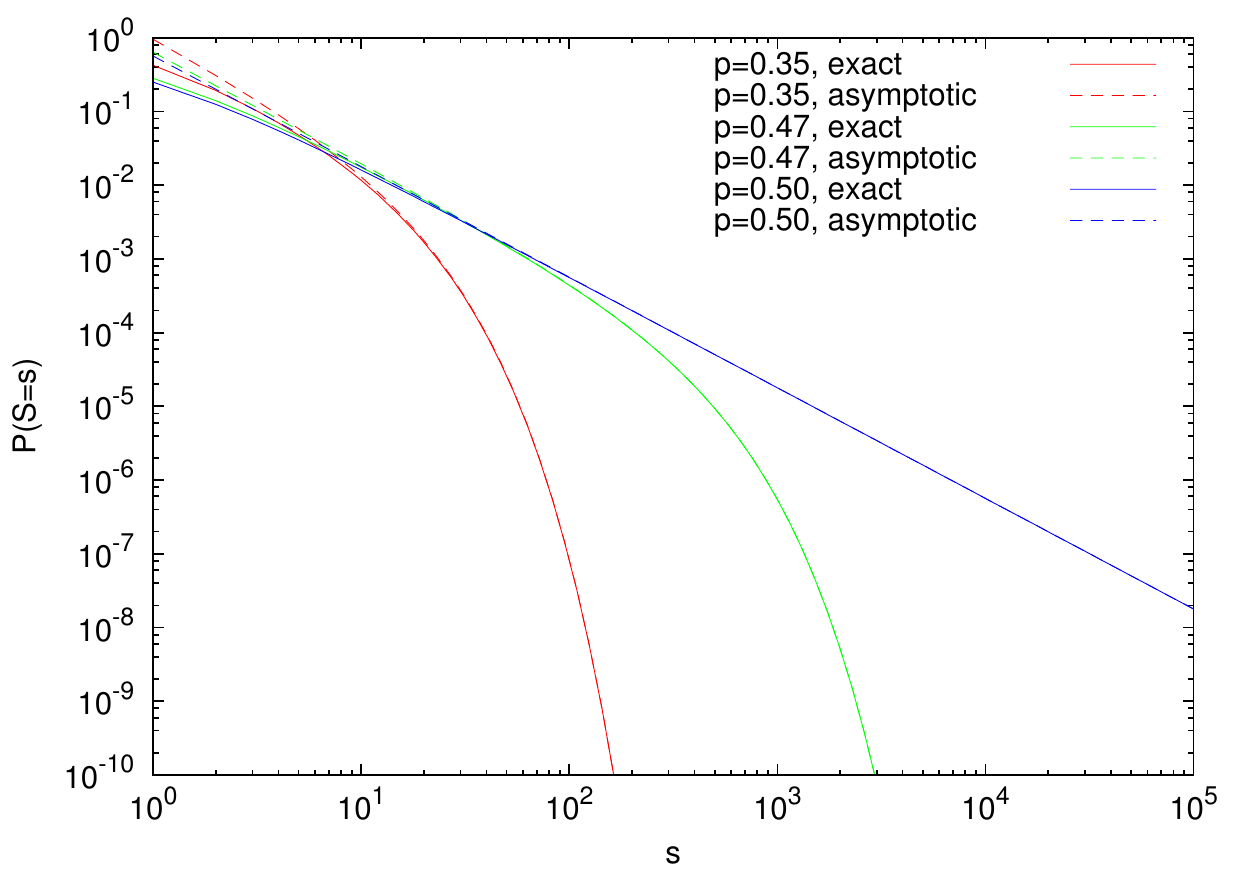}
	\caption{Probability mass functions of the total size of the population $S$ ,
for different values of the parameter $p$ of a binomial distribution with $n=2$, both in the subcritical and critical cases.
The asymptotic solution for large $s$ is also shown.
The pure power law at the critical point becomes apparent. }
	\label{tail}
\end{figure}

Nevertheless, the exact expression we have obtained for $P(S=s)$ does
not teach us anything about the behavior of this function
(unless one has a great intuition about the behavior of the binomial coefficients).
In this regard, Stirling's approximation is of great help \citep{Christensen_Moloney}.
It states that, in the limit of large $N$ one can make
the substitution
\begin{equation}
N! \sim \sqrt{2\pi N} \left(\frac N e\right)^N,
\end{equation}
see the Appendix once more.
The symbol $e$ is nothing else than the $e$ number.
So, for large sizes we can apply the approximation to $s$
and also to $2s$,
\begin{equation}
(2s)! \sim \sqrt{4\pi s} \left(\frac{2s} e\right)^{2s}.
\end{equation}
Therefore, the binomial coefficient turns out to be, 
\begin{equation}
\left(
\begin{array}{c}
2s \\ s \\
\end{array}
\right)
=\frac {(2s)!}{s!s!}
\sim \frac{1}{\sqrt{\pi s}} \frac {(2s)^{2s}}{s^{2s}}
\sim \frac{4^{s}}{\sqrt{\pi s}}, 
\end{equation}
and the Catalan number, replacing $s+1 \sim s$,
\begin{equation}
C_s= \frac 1 {s+1}
\left(
\begin{array}{c}
2s \\ s \\
\end{array}
\right)
\sim \frac{4^{s}}{\sqrt{\pi} s^{3/2}}.
\end{equation}
This is an exponential increasing function of $s$,
and the term $s^{3/2}$ does not seem to play any role, asymptotically.
However, introducing the factor $p^{s-1}(1-p)^{s+1}$, we go back to equation~\eqref{catalan1}, getting
\begin{equation}
\label{catalan2}
P(S=s) \sim \frac {1-p}{\sqrt{\pi} p} \frac{[4p(1-p)]^s}{s^{3/2}}.
\end{equation}
Notice that $p(1-p)$ is no larger than $1/4$,
so the exponential term becomes decreasing, 
except for $p=1/2$, where it disappears.
We can go one step further, by writing,
\begin{equation}
[4p(1-p)]^s = e^{s \ln[4p(1-p)]} = e^{-s/\xi(p)} 
\end{equation} 
with the characteristic size defined as
\begin{equation}
\xi(p)
=\left(\ln \frac 1 { 4p(1-p)}\right)^{-1},
\end{equation}
and finally equation \eqref{catalan2} reads,
\begin{equation}
P(S=s) \sim \frac {1-p}{\sqrt{\pi} p} \frac{e^{-s/\xi(p)}}{s^{3/2}},
\end{equation}
So, for $s$ large, but substantially smaller than $\xi(p)$,
the size probability mass function is a power law, with exponent $3/2$.
For larger $s$, the exponential decay dominates.
The exception is the critical case, $p=1/2$,
for which $\xi(p)$ becomes infinite, the exponential disappears
and the distribution is a pure power law.
In this case the exponent 3/2 is a critical exponent.
The reader can see the goodness of the approximation in Fig. \ref{tail}.

Another critical exponent arises for the divergence of the characteristic
size $\xi(p)$. Introducing the deviation with respect to the critical point, 
$\Delta\equiv p-p_c = p-1/2$, one can write,
\begin{equation}
p(1-p) =\frac 1 4 - \Delta^2,
\end{equation}
and so, close to the critical point (for small $\Delta$),
\begin{equation}
\frac 1 {4p(1-p)}=\frac 1 {1-4\Delta^2} \simeq 1+4\Delta^2 + \dots
\end{equation}
(using the formula of the geometric series), then 
\begin{equation}
\ln \frac 1 {4p(1-p)} \simeq \ln(1+4\Delta^2) \simeq 4\Delta^2+\dots
\end{equation}
(using the Taylor expansion of the logarithm at point 1) and 
\begin{equation}
\xi(p)=\left(\ln \frac 1 {4p(1-p)}\right)^{-1}  \simeq \frac 1 {4\Delta^2} +\dots
\end{equation}
Therefore, the characteristic size $\xi(p)$ diverges at the critical
point as a power law, with an exponent equal to 2.
This allows to write the asymptotic formula ($s$ large) 
for the size distribution in a simpler form, 
close to the critical point ($\Delta$ small),
\begin{equation}
P(S=s) \sim \frac {1-p}{\sqrt{\pi} p} \frac{e^{-4 (p-p_c)^2 s}}{s^{3/2}}.
\end{equation}

Hence, after this perhaps long but worthwhile digression,
we are able 
to say something about the energy distribution 
in Otsuka's model,
which the reader will have already noted is a particular case of the Galton-Watson process.
If one takes $p<1/2$ the resulting energy distribution
has an exponential tail, with a characteristic scale given 
by $\xi(p)$. This means that earthquakes attenuate, or get extinct,
and in no way can dissipate energies larger than the scale 
provided by $\xi(p)$
(the probability of having an earthquake of size larger
than $10\xi(p)$ is ridiculously small).
This is the subcritical case.
On the other hand, if $p>1/2$ there are two types of earthquakes, 
first, those similar to the subcritical ones, with a size limited
by the scale defined by $\xi(p)$,
and second, infinite or never-ending earthquakes ($P_{ext} < 1$),
where the initial small perturbation
(the toppling of just one domino piece)
grows exponentially.
This is the supercritical regime \citep{Ben_Zion_review}.
Neither the subcritical nor the supercritical case
are in correspondence with the Gutenberg-Richter law,
which yields a power-law distribution of energies, 
and therefore the absence of a characteristic scale.
But this is precisely what corresponds to the critical 
case, $p=1/2$, which yields also a power-law distribution.
Thus, the propagation of an earthquake
through a fault is not only stochastic in the sense that when a patch fails one
does not know what will happen next, 
but it is worse than that, as a critical process is equally likely to intensify or attenuate.
Note how difficult is to achieve a critical
behavior, as $p$ has to be finely tuned to $1/2$, otherwise criticality is lost.
In terms of domino topplings this is what is really difficult, and not to get a
full-system supercritical toppling, which, despite its mathematical triviality, 
deserves a lot of attention from the media when a Guinness world record
is broken.

The agreement between the model and real earthquakes 
is qualitative but not quantitative, 
as the model leads to $\alpha=3/2$ whereas for earthquakes
$\alpha \simeq 5/3 \simeq 1.67$.
In the next subsection we will explain that 
the model value of 3/2 is rather robust and
other versions of the Galton-Watson process lead
to the same exponent.
This discrepancy has been explored in detail by \cite{Kagan_tectono10},
who argues that there are a series of 
technical artifacts that make increase 
the value of the exponent for earthquakes,
and therefore, following Kagan, 
both exponents would be close and probably compatible.


\subsection{Generating function of the total size of the population}

In order to advance further in the understanding of branching processes,
our little story carries us to 
the U.S. during the Second World War.
While soldiers were fighting in the field
and civilians were suffering the horrors of war, 
a group of scientists gathered in the peace of Los Alamos, New Mexico, 
to do research to develop the first nuclear bombs.
Among these brilliant people was the great Polish mathematician Stanislaw Ulam, 
who was hired by his famous colleague John Von Neumann \citep{Ulam_adventures}.
Together with David Hawkins
(philosopher of science and most talented amateur mathematician 
ever known by Ulam)
they were investigating the multiplication of neutrons
in nuclear chain reactions, using what we call now 
branching processes.
It seems that they were unaware of the pioneering work of
Galton and Watson.


Hawkins and Ulam showed, among other things, 
that the generating function $g(x)$ of the total size 
of the population, $S=\sum_{\forall t} Z_t$, fulfills, in the non-supercritical case,
\begin{equation}
g(x) = x f(g(x))
\end{equation}
where, as usual, $f(x)$ is the generating function of the number of offsprings 
of an individual element.
What follows in this subsection is based in their work for the
Manhattan Project \citep{Hawkins_Ulam,Ulam_analogies},
but our derivation is somewhat simpler.
What we call total size of the population will correspond
to all neutrons generated during the reaction.

\begin{figure}[t]
		\begin{subfigure}[b]{0.5\textwidth}
                \centering
                \includegraphics[width=\textwidth]{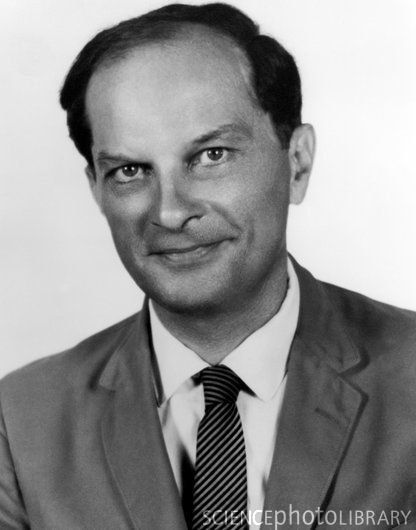}
        \end{subfigure}
        \begin{subfigure}[b]{0.5\textwidth}
                \centering
                \includegraphics[width=\textwidth]{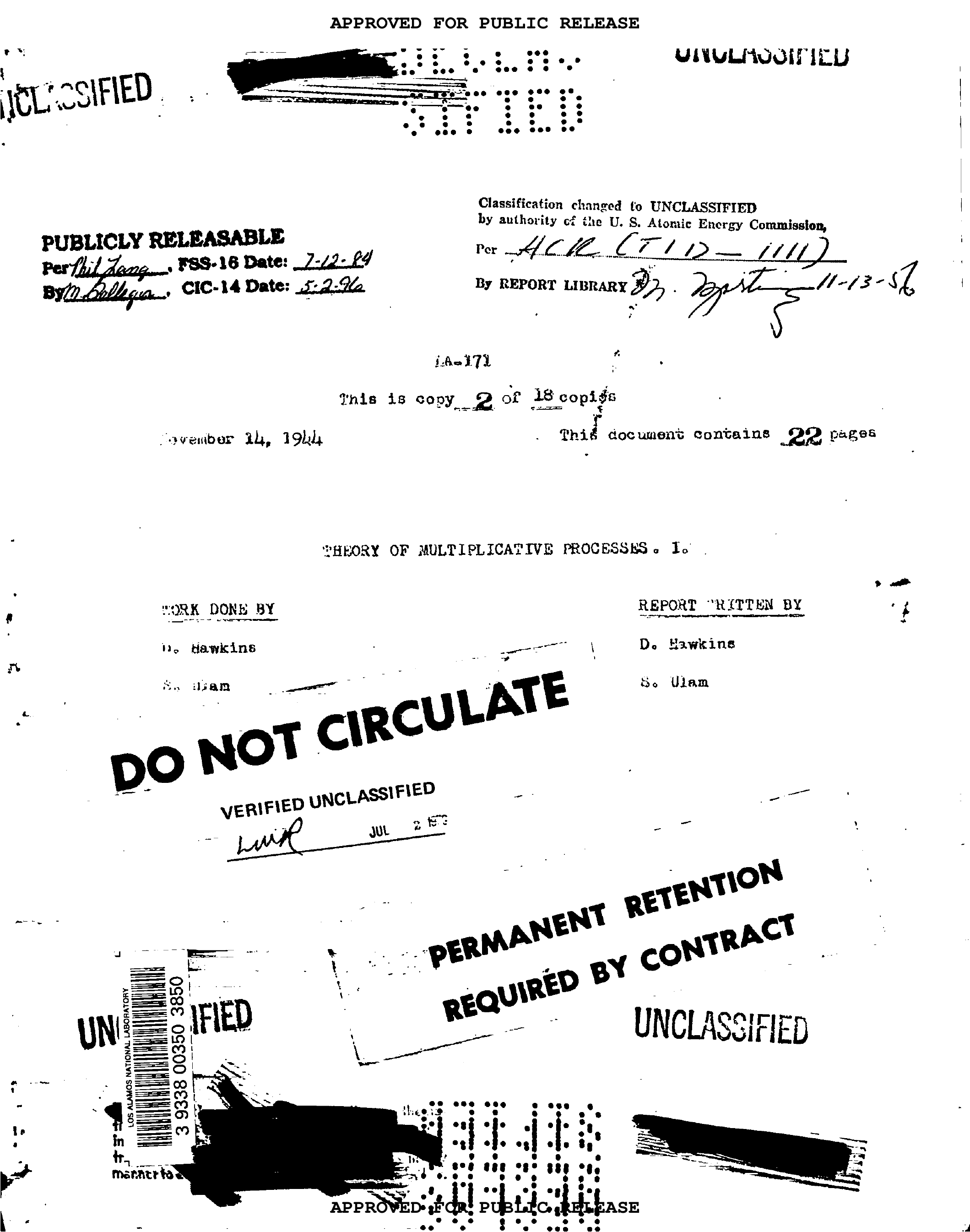}
        \end{subfigure}
\caption{Polish mathematician Stanislaw Ulam, together with the current version of the 
first page of Ref. \cite{Hawkins_Ulam}, after being unclassified for public release.
The work, in which important formulas for branching processes are derived, 
was done as a part of the Manhattan project.}
\end{figure}

First, it is convenient to consider the size from generation 
1 to $\tau$ (excluding by now the zero generation).
This is
\begin{equation}
S_{\tau}=\sum_{t=1}^{\tau} Z_t
\end{equation}
with probabilities $q^{(\tau)}_s=P(S_{\tau}=s)$ and a generating function $\tilde g_{\tau} (x)
=\sum_{\forall s} q^{(\tau)}_s x^s$.
A size $s$ in generations from 1 to $\tau$
can be decomposed into a size $k$ in the first generation, with probability $p_k$,
and a size $s-k$ in the remaining $\tau-1$ generations (from $2$ to $\tau$), 
but starting with $k$ elements; this has a probability $q^{(\tau-1,k)}_{s-k}$.
(Note that, with this notation $q^{(\tau)}_s=q^{(\tau,1)}_s$.)
Then, using the law of total probability, 
\begin{equation}
q_s^{(\tau)} = \sum_{k=1}^s p_k  q_{s-k}^{(\tau-1,k)},
\end{equation}
except for $s=0$, where $q_0^{(\tau)} = p_0$.
If we multiply by $x^s$ and sum for all $s$, from $0$ to $\infty$, 
we will obtain on the left hand side
the generating function of $S_{\tau}$, which turns out to be
\begin{equation}
\label{gtau}
\tilde
g_{\tau} (x) = 
p_0+\sum_{s=1}^\infty 
\sum_{k=1}^{s}  p_k q_{s-k}^{(\tau-1,k)} x^{s} 
=
p_0+\sum_{k=1}^\infty p_k \left[ 
\sum_{s=k}^{\infty} q_{s-k}^{(\tau-1,k)} x^{s-k}\right] x^k. 
\end{equation}
The term inside the square brackets is the generating function 
of the size from $1$ to $\tau-1$ generations but, instead of starting
with one single element (the usual $Z_0=1$),
starting with $k$ elements ($Z_1=k$).
As these $k$ parents are independent of each other, the resulting
size will be the sum of $k$ independent random variables, 
each with generating function $\tilde g _{\tau-1}(x)$,
which yields $[\tilde g _{\tau-1}(x)]^k$ as the corresponding generating function, 
that is,
\begin{equation}
[\tilde g _{\tau-1}(x)]^k = \sum_{s-k=0}^{\infty} q^{(\tau-1,k)}_{s-k} x^{s-k},
\end{equation} 
Substituting into equation~\eqref{gtau}, this leads to 
\begin{equation}
\tilde
g_{\tau} (x) = p_0+\sum_{k=1}^\infty p_k [\tilde g _{\tau-1}(x)]^k  x^k = 
f(x \tilde g _{\tau-1}(x))
\end{equation}
where we have introduced the definition of $f(x)=f_K(x)$.

If we want to include the zero generation in the size, 
we need to add an independent variable with generating function $x$
(as $Z_0$ takes the value 1 with probability 1),
and then, the generating function of the size from generation $0$ to $\tau$ 
is the product $g_{\tau} (x)= x \tilde g_{\tau} (x)$.
This leads to 
\begin{equation}
g_{\tau} (x) = x f( g _{\tau-1}(x)).
\end{equation}

Coming back to the total size,
\begin{equation}
S= \sum_{t=0}^\infty  Z_t,
\end{equation}
the corresponding generating function is $g(x)=\lim_{\tau \rightarrow \infty} g_{\tau}(x)$.
If the probability of extinction is one, i.e., if the system is not supercritical,
this is the same as $\lim_{\tau \rightarrow \infty} g_{\tau-1}(x)$, 
and therefore we have
\begin{equation}
\label{recur}
g(x)=xf(g(x)).
\end{equation}
So, the desired generating function is the solution of this equation, 
with $f(x)$ known.
We will not be able to solve it in general;
however, notice  that this is not necessary in order to get
the moments of $S$.
Differentiating equation~\eqref{recur} with respect $x$ one obtains
\begin{equation}
g'(x)=f(g(x)) + xf'(g(x))g'(x),
\end{equation}
and taking $x=1$ 
and isolating,
\begin{equation}
\langle S \rangle = g'(1) = \frac 1 {1-f'(1)}=\frac 1 {1-m},
\end{equation}
which goes to infinity as $\langle K \rangle =m=f'(1)$ goes to 1, 
that is, at the critical point.
Of course, as we have mentioned, the result is not applicable
in the supercritical case, $m >1$,
where the population can growth to infinite with a non-zero probability.
Further differentiation yields higher-order moments.

The same result could have been obtained directly, as 
\begin{equation}
\langle S \rangle = \langle Z_0+Z_1+Z_2 + \cdots\rangle
=\langle Z_0\rangle +\langle Z_1\rangle+\langle Z_2\rangle + \cdots
=1+m+m^2+\cdots =\frac 1 {1-m},
\end{equation}
where the last equality only holds in the subcritical case, 
otherwise, $\langle S \rangle$ goes to infinity.

In a few cases, the equation for $g(x)$ allows
to easily obtain  a solution.
Revisiting the binomial example with $n=2$, for which 
$f(x)=(1-p+p x)^2$, one gets
\begin{equation}
g(x)=xf(g(x)) = x (1-p+p g(x))^2,
\end{equation}
from where 
\begin{equation}
g(x)=\frac{1-2pqx \pm \sqrt{1-4pqx}}{2p^2x},
\end{equation}
with $q=1-p$.
Using the Taylor expansion for the square root term (see the Appendix),
\begin{equation}
\sqrt{1-4pqx} = 1-2pqx-\sum_{s=1}^\infty\frac{(2s-1)!!2^{s+1}}{(s+1)!} (pqx)^{s+1},
\end{equation}
and recognizing the Catalan numbers $C_s$ there, we get (see the Appendix),
\begin{equation}
g(x) = \frac q p \sum_{s=1}^\infty C_s(pqx)^s,
\end{equation}
where we also realize that only the minus sign before the square root
leads to a true generating function.
Therefore, the coefficients of $x^s$ lead to
\begin{equation}
P(S=s) = C_s p^{s-1} q^{s+1},   
\end{equation}
for $s\ge 1$. This result is exactly the same
as the one we obtained previously in a different manner (see equation~\eqref{catalan1}),
although in this way we do not need to count trees,
as the Catalan numbers arise directly in the series expansion
(in fact, we do not even need to know them).

We confirm that the results for Otsuka's binomial model 
yield a size exponent equal to 3/2.
But it would be desirable to test the robustness of such exponent
value, as, after all, the model is a crude simplification of 
reality, and we would like that modifications of the model
do not lead to a totally different behavior.
Despite the difficulty to find the power-law behavior
(for which we need to finely tune the parameter $p$ to 1/2),
if one considers other models different than the binomial one, 
the asymptotic behavior of the size distribution
is in general always given by a power law with exponent 3/2, in the critical case; 
this can be proved by means of Cauchy's formula and assuming only finite variance, 
see \cite{Otter_1949,Harris_original}. 
So, going beyond robustness, it is common to denote such invariance as universality.

\subsection{Self-organized branching process}

At this point we are ready to accept
the agreement, not only qualitative but, 
following Kagan's remarks \citep{Kagan_tectono10}, also quantitative,
 between a critical branching process 
and earthquake occurrence.
So, in order to tune the model to reality
we just need to take $p=1/2$ (in Otsuka's binomial case)
or $m=1$ (in general) and the agreement is really satisfactory,
and we could finish our search for a model here.

But we can try to go one step farther and ask:
why do we find that the tectonic systems 
(and other geosystems related to natural catastrophes)
are always
keeping a delicate balance between 
a subcritical and a supercritical state, 
i.e., in an apparent critical state? Can the coincidence be just fortuitous?
In the reproduction of individuals 
one could devise an evolutionary explanation.
Imagine a series of isolated islands, each one occupied by a population
following a Galton-Watson process but with different parameters
for each island.
It is clear that islands with subcritical populations get deserted
after a number of generations.
Populations in supercritical islands either get extinct also
or explode exponentially,  in which case we assume that 
the population collapses, due to the exhaustion of the resources
(this is an ingredient that is not in the original Galton-Watson model).
In the critical case, the population also gets extinct, 
but for a few of these islands the population can survive for 
very long times, much longer than in the subcritical and supercritical cases.
So, after a long enough time we would only find critical populations.

However, this evolutionary scenario is not applicable to a tectonic
system, where, when the process (the earthquake) gets extinct, 
a new one will start sooner or later. 
Rather, 
the situation would be analogous to finding all magnetic materials on Earth
at the onset of magnetization, 
which would mean that their temperatures would be equal to the 
Curie temperature of each material. 
One could suspect then that there is some mechanism enforcing
criticality, where the temperature changes as a function of magnetization,
and magnetization is kept at the border of the transition;
in other words, both parameters are linked through some feedback mechanism
\citep{Sornette_1992,Pruessner_Peters_pre}.

\cite{Zapperi_branching} propose a model in this line.
They start with a standard branching process but introduce some important
modifications:
\begin{itemize}
\item They limit the number of generations to a maximum $\tau$, so $0\le t\le \tau$.
\item After the extinction of the process 
(which is obviously certain when the number of generations is limited),
the parameters of the process change for the next realization, in
such a way that for subcritical cases ($m<1$), the mean $m$ of the number
of offsprings for each individual unit increases, 
whereas in the supercritical case ($m>1$) the mean $m$ decreases.
The idea is to make the critical state $m=1$ an attractor of the dynamics.
\end{itemize}

In order to be more concrete, let us consider the usual binomial
distribution with only 0, 1, or 2 possible offsprings
and a probability $p$ that each reproductive trial is successful.
Then we already know that $p<1/2$, $p=1/2$, and $p>1/2$
correspond to the subcritical, critical, and supercritical cases, 
respectively.
The dynamics proposed by Zapperi and coauthors relies on the activity that 
reaches the ``boundary'' of the system (defined by the last generation, $t=\tau$),
which is $Z_\tau$, changing the probability $p$ through the following formula
\begin{equation}
\label{pT1}
p(T+1) = p(T) + \frac{1-Z_\tau(p(T),T)}{N},
\end{equation}
with 
$T$ a discrete time index counting the number 
of realizations of the process (do not confuse with $t$) and
$N=2^{\tau+1}-1$ the maximum number of possible elements,
i.e., the number of branches of the underlying complete tree.
Thus, if the activity does not reach the boundary,
$Z_\tau$ is zero and
the parameter $p$ is increased by $1/N$,
this is a very small number in the limit of very large systems 
($N\rightarrow \infty$).
On the other hand, if the activity at the boundary is 
greater than one, $p$ is decreased by $(Z_\tau-1)/N$. 

We already know that the expected value of $Z_\tau$ is $m^\tau$,
with $m$ the mean of the offspring distribution ($m=2p$ in our particular binomial model).
Let us introduce a noise term, $\eta$, which takes into account the fluctuations
of $Z_\tau$ with respect its mean, i.e., $\eta=Z_\tau-m^\tau$.
Obviously, by construction, $\langle \eta \rangle =0$.
If we neglect, for a while, the noise term in equation \eqref{pT1}, the deterministic part reads,
\begin{equation}
p(T+1) = F(p(T)) = p(T) + \frac{1-(2p(T))^\tau}{N}.
\end{equation}
This is a discrete dynamical system, or a map, 
for which a fixed point $p^*=F(p^*)$ exists,  $p^*=1/2$.
Moreover, the fixed point is attractive, as
$|F'(p^*)| <1$ \citep{Alligood_chaos}, due to $\tau \ll N$.

Taking into account the value of the standard deviation
of $Z_\tau$ \citep{Harris_original}, it can be shown that
the noise term $\eta/N$ will have a vanishing effect
in the limit of very large systems, and then the stochastic evolution
will lead the system towards the deterministic fixed point, plus 
small random fluctuations around it.

This spontaneous evolution of a system towards a particular
organized state is referred to as self-organization.
It is clear now that what Zapperi et al. introduced 
is a branching process that self-organizes towards a critical state.
Nevertheless, the particular dynamics they propose 
seems a bit arbitrary.
How can this kind of global control be implemented in a real system, 
where we expect the interactions between elements to be purely local?

\subsection{Self-organized criticality and sandpile models}

In fact, 
the self-organized branching process introduced
by \cite{Zapperi_branching} was naturally embedded
in the previous notion of self-organized criticality (SOC), 
invented by Bak and coworkers in the 1980's \citep{Bak_book,Jensen,Christensen_Moloney}.
Although it is not relevant for our story, 
it is worth to state that these authors were not interested in
(because they were not aware of)
the problem of power-law distributions in natural hazards \citep{Bak_book};
rather, they were mainly concerned to similar-in-spirit problems
in condensed-matter physics, as charge density waves and 
one-over-f noise, as well as to the emergence of fractal spatial structures elsewhere
\citep{BTW87}.
The fact that earthquakes (and other hazards) were a manifestation
of self-organized criticality was a fortunate by-product, pointed 
by 
\cite{Ito_Matsuzaki}, \cite{Sornette_Sornette}, and \cite{Bak_Tang}
shortly after the introduction of the SOC concept,
see also the review of \cite{Main}. 
Nowadays, natural hazards are one of the main applications
of SOC, despite the original lack of attention by \cite{BTW87}.
As we have seen through this chapter, 
ignorance seems a common characteristic of science evolution.

The metaphor used by Bak in order to illustrate his ideas
was that of a pile of sand \citep{Bak_book}.
We have to recognize that the sandpile we are going to consider
is a bit esoteric; in fact, there is a clear correspondence
between the model and a pile only in one dimension
(the one-dimensional model corresponds to a pile constrained in two dimensions,
between two parallel plates \citep{Christensen_oslo}).
But instead of keeping close to reality, 
it is more effective to deal with a mean-field sandpile;
this is achieved either in a system defined in the limit of 
infinite dimensions or in a system in which each element
has ``random neighbors'',
and neglecting the correlations between the elements.
Notice that Bak and colleagues make use of a new concept, 
not present in the branching processes already explained: 
the notion of complexity, understood here as the nontrivial interaction between many units or agents,
which will result in an emergent collective behavior 
that is different than the sum of the behavior of the individual parts
\citep{Newman_survey}.

So, consider a system consisting in a large number of elements, such that each element can store
a certain number of discrete packages (or particles), but when this limit is surpassed
the packages are released to other elements -- the neighbors.
The situation is analogous to what happens 
in a Ministry office. Each bureaucrat has a series of documents or papers
(the packages) at his/her desk, but when the number
of those is too big, he/she decides to do something about it and transfers 
some papers to some other (random) bureaucrats, and so on \citep{Bak_book}.
This simple behavior will lead to interesting dynamics, unexpectedly.

To be specific, let us consider that
each element can store at most one package;
if some extra package arrives to it, the element releases
two packages to some other units, taken randomly 
(either among all other elements, what defines random neighbors
or among the $2d$ nearest neighbors in a $d-$dimensional square lattice).
If, after the release, the number of packages is still greater than one
(which may happen if the element received more than one package)
the release process is repeated.
All the elements evolve following a
parallel updating of their dynamics, 
i.e., there is a common clock 
setting the time $t$ of all elements.
In a formula,
\begin{equation}
\mbox{if } z_i \ge 2 \Rightarrow \left \{
\begin{array}{lll}
z_{n(i)} & \rightarrow & z_{n(i)} + 1,\\
z_i & \rightarrow & z_i -2,\\
\end{array}
\right.
\end{equation}
where $z_i$ counts the number of packages of element $i$
and $n(i)$ denotes two of its neighbors.

Obviously, this process can give rise to an avalanche in the transference of
packages, which only stops when all elements
have no more than one package.
In that case, the system is perturbed by the addition
of one extra package to a randomly chosen element,
and the dynamics starts again. This defines a new
time scale, denoted by $T$ (in the same way as in the previous subsection).
So, 
\begin{equation}
\mbox{if } z_i \le 1, \forall i \Rightarrow 
\begin{array}{lll}
z_{j} & \rightarrow & z_{j} + 1,\\
\end{array}
\end{equation}
where $j$ denotes a randomly selected unit.
The system also releases packages outside
(or to the garbage can, in the bureaucrats picture);
in a $d-$ dimensional lattice this happens when a boundary
element selects as a neighbor an external element;
in a fully random-neighbor system
this happen just with a small predefined probability
for each element.
This simple variation of the original sandpile model 
of \cite{BTW87} (changing the topology of the system by means 
of a different selection of neighbors)
can be viewed also as a mean-field version of the so-called
Manna model \citep{Manna,Christensen_Moloney}.

The simple rules of the model make that the total number of packages in the system, $M$,
evolves, from the addition of one package to the next, accordingly to
\begin{equation}
\label{MT1}
M(T+1)=M(T)+1-\mbox{drop}(T),
\end{equation}
where drop is the number of packages that 
are expelled from the system.
The key parameter of this model is $p$,
defined, for each element, as the probability that its number
of packages is equal to one
(so they are at the onset of instability).
But in a mean field description all elements are uncorrelated and equivalent, 
so we can define a generic $p$ for the whole system, 
verifying $p=M/N$, with $N$ the total number of elements.
So, there is a probability $p$ that an element releases two packages
when it receives one.
The action of release is what constitutes the generation of an offspring, 
which is the element that relaxes.
Therefore, dividing equation~\eqref{MT1} by $N$
we obtain
\begin{equation}
p(T+1)=p(T) +\frac {1-\mbox{drop}(T)} N,
\end{equation}
which we can recognize as essentially equation \eqref{pT1}, the one introduced by \cite{Zapperi_branching} 
in the self-organized branching process.
We have already realized that this equation provides
a feedback mechanism of the number of packages into
the toppling (branching) probability
(early identifications of this obvious feedback in SOC
were written by \cite{Kadanoff91} and \cite{Sornette_1992}).

Both in the limit of an infinite dimension lattice 
or in a fully random neighbor system one realizes that 
the evolution of an avalanche corresponds to a set of propagating
non-interacting packages (as the probability that the activity
comes back to an element is vanishingly small), and therefore the activity evolves as
a branching process. But note that the tree associated to the branching process
does not correspond to a quenched underlying structure of the system, 
as the random neighbors are selected dynamically, at each time step.
The limit $\tau$ in the number of generations introduced by Zapperi and coauthors
needs to be added as an extra ingredient in the model,
enforcing the dissipation of packages to take place at
the $\tau$ time step.
In summary, this illustrates the correspondence between 
the mean-field limit of sandpile models and branching processes.
This is enough for our purposes.
Other chapters in this book illustrate in much more detail
the dynamics of sandpiles.
Nevertheless, it is worth mentioning
that the first connection between
SOC and critical branching process was
published by \cite{Alstrom},
where it was assumed, however, that the system was in a critical state from the beginning.
Notably, much before, \cite{Vere_Jones76} had proposed a branching model very similar to 
Otsuka's (but, as usual, unaware of it) and realized that the tectonic system
should evolve spontaneously towards criticality.
Also, very recently, \cite{Hergarten12} has introduced a variation of Zapperi et al.'s
branching model that evolves only with local rules.

Recapitulating, self-organized criticality offers a coherent framework
for the understanding of earthquakes
and many other natural hazards mentioned in the first section. 
Indeed, both phenomena (SOC and earthquakes)
show a highly non-linear response, 
where a small and slow perturbation or driving
(the addition of grains, or the stress provided by the motion
of the tectonic plates) pumps energy into the system, 
which, due to the presence of local thresholds stores that energy,
until at some point some threshold is surpassed.
The resulting release of energy propagates locally, which can
trigger further surpassings of thresholds, generating a chain reaction
or avalanche.
One key point is that the energy released in such a way has to be
power-law distributed, so the system responds in all possible scales.
Notice also that the dynamics shows a time-scale separation, 
as the avalanches happen infinitely fast compared with the driving
(the toppling of grains is stopped during the propagation of an avalanche).
Moreover,
\cite{Main} mentions additional characteristics of seismicity
present in SOC models, namely, 
stress drops that are small in comparison
with the regional tectonic stress field 
and the existence of
seismicity induced or triggered by
relatively small stress perturbations.
All this makes SOC a very plausible mechanism for earthquakes.
The connection is made still more concrete using variations of the sandpile models
that mimic the behavior of the spring-block model of \cite{Burridge_Knopoff}
as the so-called OFC model \citep{OFC}.
See also \cite{Main}.

However, as far as we know, the authentic hallmark of SOC, 
the existence of an underlying second-order (continuous) phase
transition, has not been found in earthquakes.
The very nature of SOC makes almost impossible to identify
such an abrupt change of an order parameter when a control parameter
changes (because the control parameter is attracted towards the critical point).
Nevertheless, this elusive behavior has been found in a different
system: rainfall \citep{Peters_np}, thanks to very large fluctuations
from criticality;
so, if a control and an order parameter could be measured
and if similarly large fluctuations were exist,
one would finally prove the existence of SOC in earthquakes.

The same reasoning applies to other natural hazards,
for which, at least, sandpile-like models are abundant
in the literature, and their classification as SOC systems
is plausible \citep{Jensen}.
The case of hurricanes is still not clear \citep{Corral_Elsner},
whereas for tsunamis we can state that their power-law distribution
\citep{Burroughs} does not
arise from a SOC mechanism, as they are not slowly driven
(rather, they are violently driven by earthquakes, landslides and
meteorite impacts).

Finally, it is worth mentioning that there is another connection between 
branching processes and earthquakes.
Instead of using the branching to model the propagation of individual 
earthquakes, it is used for the way in which one earthquake
triggers other earthquakes, i.e., aftershocks,
following the so-called Omori law.
The most representative model of this kind
is the epidemic-type aftershock-sequences (ETAS) model 
\citep{Ogata99,Helmstetter_Sornette_jgr02}.
Interestingly, the evolution model of \cite{Bak_Sneppen}
(another paradigm of SOC)
can be interpreted to reproduce the statistics of earthquakes
from this (slow) time scale \citep{Ito}.
This perspective opened a whole new line in statistical seismology,
but this is a different story
\citep{Bak.2002,Corral_prl.2004,Corral_physA.2004}.

\section{Conclusions}

We started this chapter showing some remarkable statistical properties of earthquake occurrence, and ended up mingling with infinite-dimensional sandpiles models for self-organized criticality. In between, we learnt a few things about branching processes. 
Now 
we sketch some consequences for our initial object of study: natural hazards.

First, besides any model, we can say a few things just by looking at the data: earthquakes and other natural hazards follow a power-law distribution of sizes, in some cases with an exponential cutoff due to finite-size effects (the Earth is finite, after all!). For the particular values of the exponents found, this implies that, although big events are less likely, they are always the main contributors of the overall devastation. 
As financial data of asset returns and other social and technological data have also been reported to follow power law distributions 
\citep{Mantegna_Stanley,Newman_05}, 
one wonders what the points in common with these systems and natural hazards can be.


Regarding Otsuka's rupture model, we showed how, by using a fairly simple stochastic cascade 
setup for the local dynamics of fault patches and the mathematical formalism for branching 
processes, one can reproduce the global statistical properties of real earthquake occurrences 
(and other natural hazards). This is quite remarkable, as it constitutes a link between two 
distinct observational scales: the micro-scale of local dynamics, and the macro-scale of 
global statistical behavior. 


But Otsuka's model is a particular case of the Galton-Watson branching process.
So, first, we presented in an easy way the main results already known for such processes
(main results in relation to our interests). We explained how the machinery of probability generating functions allows to find a formula for the activity (or population) at any generation of the process. In the limit of infinite generations, one gets the probability of extinction, which shows an abrupt change between two different regimes: extinction for sure if the mean number of
offsprings is below or equal to one, and the possibility of non-extinction in the opposite case. Further progress leads to an expression for the probability of the total size of the process (the total population ever born or the total energy radiated by an earthquake). It is precisely at the border of the two mentioned cases, 
at the critical point of the transition, that one finds a behavior compatible
with earthquakes and other natural hazards. A power-law distribution with exponent 3/2 emerges in this case; however, 
it remained unexplained how the Earth should drive itself towards such a critical state.

In this regard, we showed how, by using a simple feedback mechanism, one can turn the critical point into an attractor of the model. A global condition, related with boundary dissipation, acts on the probability of activation,
in such a way that when this probability is low, it increases, 
and vice versa when it is high. Idealized sandpile models in the mean-field limit implement in a natural way this mechanism, 
by means of the transport of particles through the system up to the boundaries where they are dissipated.
The content of particles regulates the activity in the system.

It is worth mentioning that going beyond the mean-field limit and turn to lattice (more realistic) systems makes things
terribly complicated, and the researcher has to rely more and more on computer 
simulations and losses the guide of exact, or at least approximated analytical treatments.
But this makes the mathematical problems that these systems pose much more interesting and exciting. For sure, researchers will devote their efforts to them for decades.

As a final point, we have to recognize that criticality and self-organized 
criticality are not the only ways to generate power-law distributions.
In fact, much simpler processes that yield power laws exist, 
as reviewed in \cite{Sornette_critical_book,Mitz,Newman_05}.
A well known mechanism that escapes from the normal-distribution attractor
in diffusion processes is provided by anomalous diffusion \citep{Bouchaud_Georges},
and its relation with sandpiles was studied by \cite{Boguna_Corral}, among others.
Nevertheless, we believe the present work has clearly shown the plausibility
of self-organized criticality for the explanation of earthquakes and
natural hazards in general.
A complementary, even more complex perspective is provided by \cite{Ben_Zion_review}.

\section*{Appendix}
\addcontentsline{toc}{section}{Appendix}

\subsection*{Properties of power-law distributions}
\begin{figure}
\centering
\includegraphics[width=0.8\textwidth]{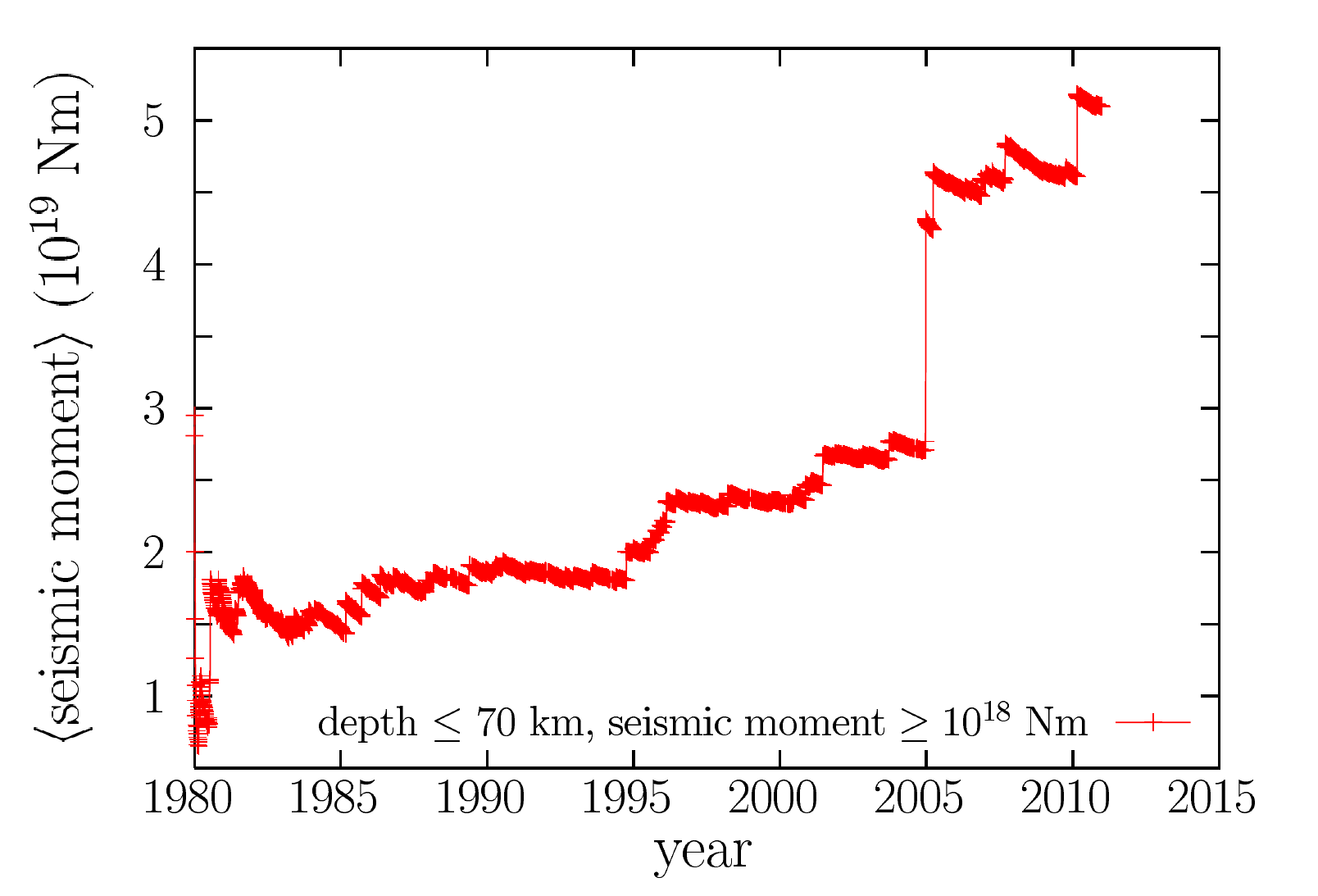}
\caption{Mean seismic moment for worldwide shallow earthquakes
with seismic moment greater that $10^{18}$ Nm,
using the CMT catalog, starting in 1980.
This yields a total of 3363 events.
Note that the mean value does not converge.
The big jump at the end of 2004 is caused
by the great Sumatra-Andaman earthquake.
The radiated energy should lead to
the same behavior.}
\label{meanEnoconv}
\end{figure}

Some facts about the power-law distribution are remarkable.
Let us consider the probability density $D(E) \propto 1/E^\alpha$,
defined between $E_{min}$ and $\infty$.
We may first calculate its mean, i.e., the expected
value of $E$, given by 
\begin{equation}
\langle E\rangle = \int_{E_{min}}^\infty E D(E) dE.
\end{equation}
It is easy to check that,
when $\alpha \le 2$ (i.e. $b \le 3/2$),
this integral becomes infinite, so, 
mathematicians would state
the expected value of the energy does not exist, 
whereas physicists would say
that that value is infinite.
We take the second option,
which is more informative
as we are aware of what we are dealing with.
Of course, the average energy radiated by an earthquake
cannot be infinite
(the Earth contains a finite amount of energy), 
so there is a problem extrapolating
the power law up to infinity.
With a normal distribution or with an exponential distribution (for example)
we would not have such a problem of extrapolation,
but it is worth to realize that this is a physical problem, 
not a mathematical problem --
for instance, 
if instead of energy we were talking about time between some events,
the mean time could perfectly be ``infinite''.
Then, for physical reasons, there has to be an upper limit for the validity of
the Gutenberg-Richter law; however, we have no idea 
about how large that limit should be.
In practice, the fact that the mean energy becomes infinite
means that the average energy one might calculate
from a series of data does not converge,
no matter the number of data. Figure~\ref{meanEnoconv} illustrates this fact
for the case of mean seismic moment, which is considered to be proportional 
to radiated energy.
Summarizing,
seismologists are totally ignorant about the 
mean energy radiated by earthquakes,
due to the special properties of power-law
distributions.

Although previously we interpreted as good news the fact that most earthquakes are
of small size and only very few of them are devastating,
the situation is certainly not so favorable.
The reason is that 
the rare big events, despite their scarcity, are the ones responsible
for the dissipation of energy in the system.
For the particular value of $\alpha$ we are dealing with,
it is easy to check that the largest order of magnitude  considered in the energy
(the largest decade, or scale)
contributes to the total budget more than all the other scales below. 
In mathematical terms,
\begin{equation}
\int _{E_{min}}^{c} E D(E) dE
<
\int _{c}^{10 c} E D(E) dE,
\end{equation}
no matter how big is $c$ (see next subsection for details).

A second peculiar property of power laws is scale invariance.
Let us introduce the concept of scale transformation,
considering an arbitrary function that we call $D(E)$.
The idea of a scale transformation is to look at the function
$D(E)$ at a different scale, as for instance, using a mathematical microscope.
We can have a view of the function at the scale of meters
(if $E$ and $D(E)$ were distances)
and try to see how it looks at the scale of centimeters.
This is performed through a scale transformation,
denoted by an operator $T$ acting on the function $D(E)$, as
\begin{equation}
T[D(E)]={c_2}{D(E /c_1)},
\end{equation}
where $c_1$ and $c_2$ are two constants called
scale parameters, performing a linear transformation
on $E$ and $D$. 
In the case of the meters-centimeters example, 
$c_1=c_2=100$.

\begin{figure}
		\begin{subfigure}[b]{0.5\textwidth}
                \centering
                \includegraphics[width=0.98\textwidth]{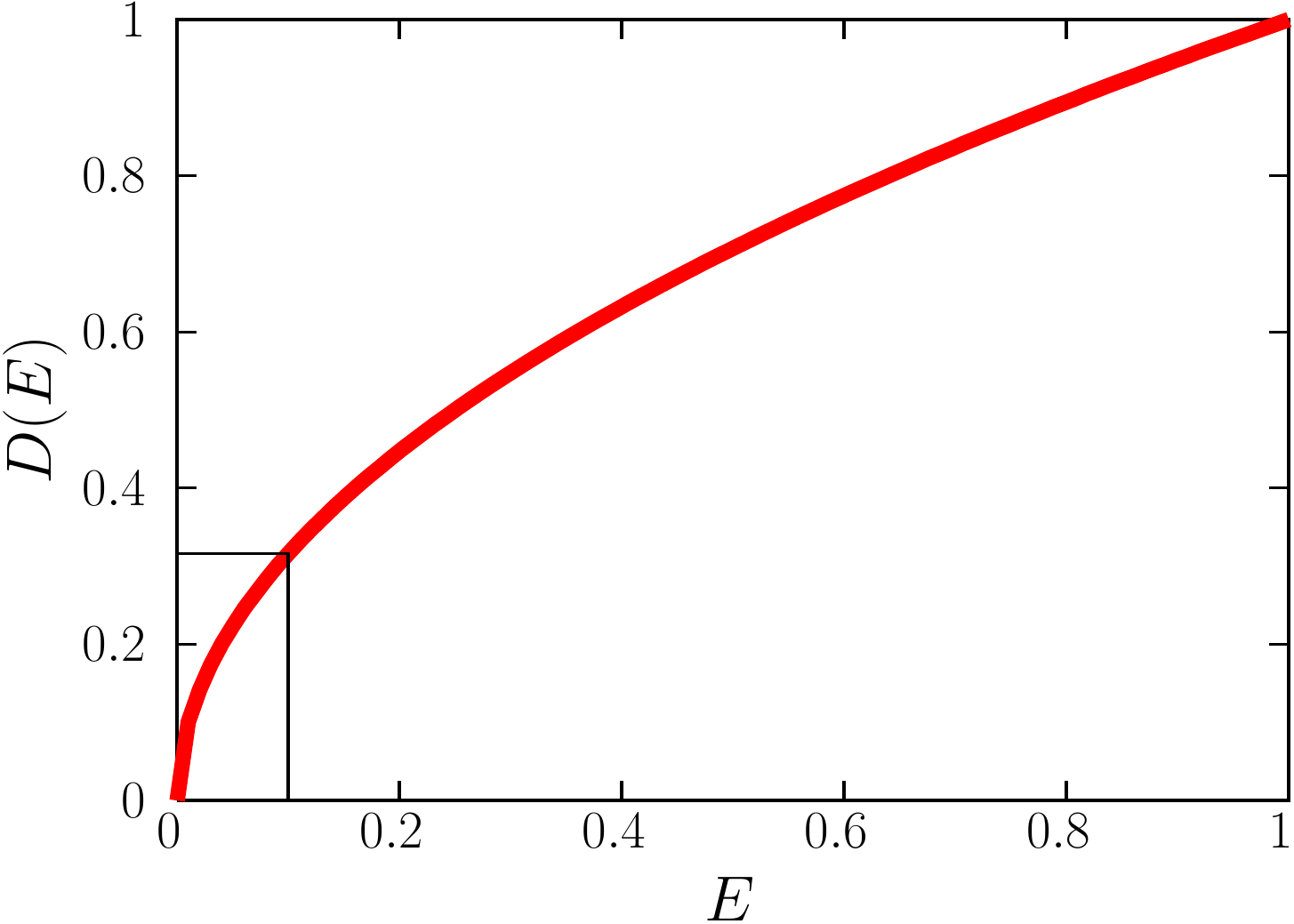}
        \end{subfigure}
        \begin{subfigure}[b]{0.5\textwidth}
                \centering
                \includegraphics[width=0.98\textwidth]{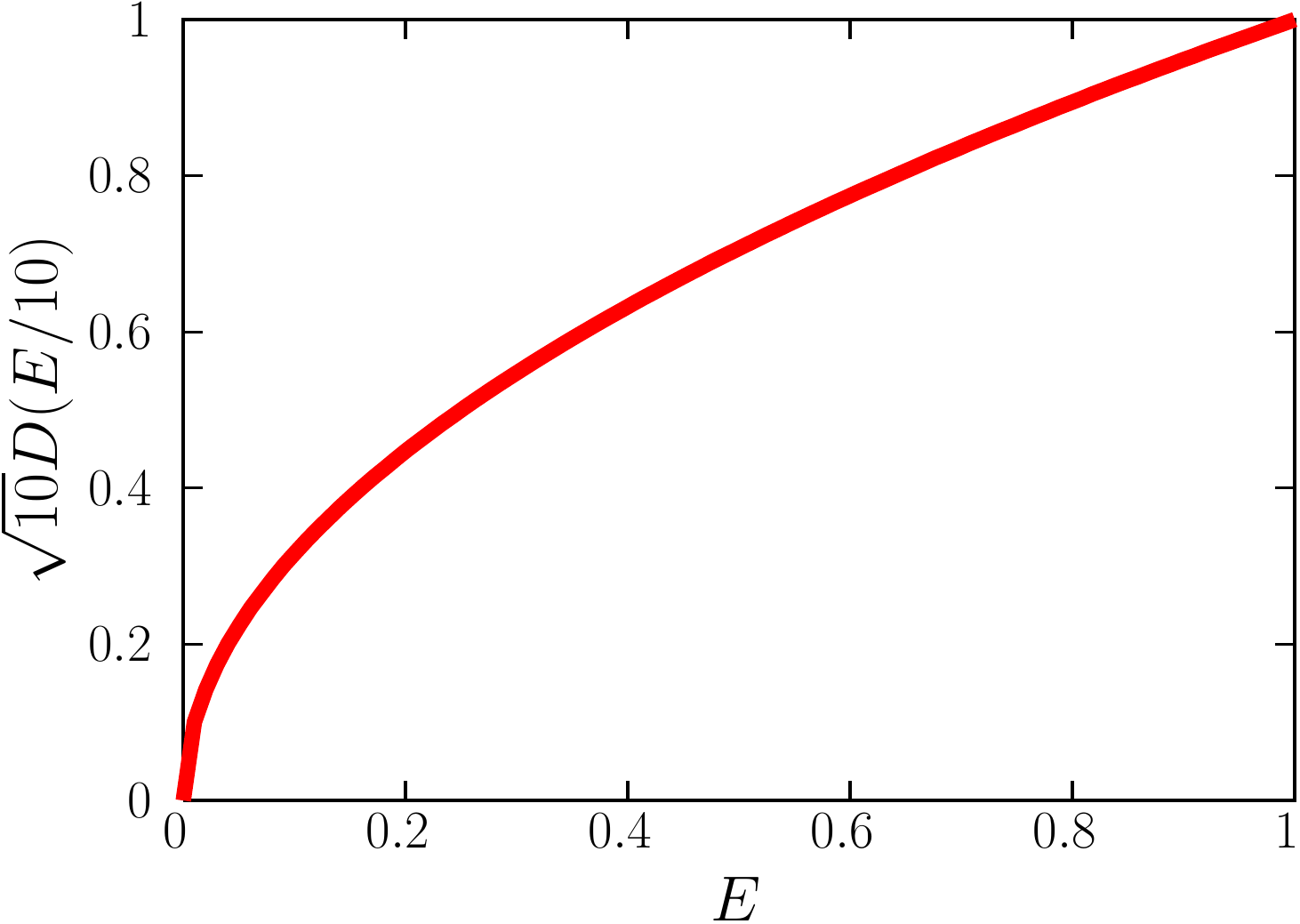}
        \end{subfigure}
\caption{A scale transformation acting on its corresponding scale-invariant function.
The function is expanded by factors $c_1=10$ and $c_2=\sqrt{10}$, in such a way that the
small box at the left is the full figure at the right. The function is $D(E)=\sqrt{E}$.}
\label{scale invariance}
\end{figure}

In general, almost every function changes
under a scale transformation;
the exception can be found looking for the 
function or functions that verify the following condition, 
\begin{equation}
D(E)={c_2}{D(E/c_1)}.
\end{equation}
It is trivial to check that a solution 
is given by the power-law function
\begin{equation}
\label{power}
D(E) \propto \frac 1 {E^\alpha}
\end{equation}
with $\alpha$ given by
\begin{equation}
\alpha=-\frac{\ln c_2}{\ln c_1},
\end{equation}
in other words, a power law with exponent 
$\alpha$ does not change under a scale transformation if
the scale factors are related through
\begin{equation}
\label{c1c2}
c_2=\frac 1 {c_1^\alpha}
\end{equation}
Figure \ref{scale invariance} shows how indeed this is the case, 
with $c_1=10$, $c_2=\sqrt{10}$, and $D(E)=\sqrt E$.
Note that the constant of proportionality in equation~\eqref{power}, contained in the symbol 
$\propto$, does not play any role here.

{
More importantly, it can also be demonstrated that not only
the power law is a solution,
but it is the only solution valid for all values of $c_1$ 
(positive real)
if $c_1$ and $c_2$ are related 
by equation \eqref{c1c2} \citep{Takayasu,Newman_05,Christensen_Moloney,Corral_Lacidogna}. 
In summary, the condition of scale invariance
demands that
\begin{equation}
D(E)={c_2}{D(E/c_1)} \, \mbox{ for all } c_1 \, \mbox{ positive real},
\end{equation}
and then, the only solution is the power law.
One can verify that other solutions, as $D(E)=\sin (\ln E)$,
only work for special values of $c_1$ and $c_2$.
}

Scale invariance is in fact the symmetry
associated to scale transformations, 
in an analogous way as rotational invariance
is the symmetry corresponding to rotations.
If scale invariance is fulfilled, 
no characteristic scale can be defined 
for the variable $E$,
in the same way as if there is rotational
invariance in a system, 
this system cannot be used to point at 
a particular direction
(a compass cannot be built from a ball).
Systems do not displaying scale invariance allow to define characteristic scales, 
as the exponential functions defining radioactive decay lead to the definition of the unit of time
in terms of the half-life.

There is, nevertheless, an important point to be taken
into account here.
If $D(E)$ represents a probability density
(as it is the case for the energy radiated by earthquakes),
then, $D(E)$ cannot be a power law for all $E \ge 0$,
because it could not be normalized (its integral from 0 to $\infty$
would diverge).
We have already mentioned that it is necessary to introduce a lower cutoff $E_{min}$
in order to avoid this fact.
Also, sometimes the power law cannot be extended to infinity, for physical reasons.
So, complete scale invariance is not possible for probability distributions, 
and one can have only a restricted scale invariance.
However, in the case of earthquakes, as both the lower limit
and the upper limit are not available from observations, 
scale invariance plays a genuine role.

Scale invariance in the energy of earthquakes has some counter-intuitive consequences.
Imagine that you arrive at a new country, and you are worried about earthquakes, 
and ask the people there the following question:
{\it how big are typically earthquakes here?}
Despite the innocence of such a simple question, 
due to scale invariance 
no characteristic scale for the energy can be defined and the question
has no possible answer.

\subsection*{Dissipation of energy in the largest scales}
\label{dissipation}

Let us consider a (continuous) power-law distribution, 
defined, for simplicity, between 1 and $\infty$,
with probability density,
\begin{equation}
D(E) \propto \frac 1 {E^\alpha}.
\end{equation}
We are going to see that, for a given $r > 2$ 
there exist values of $\alpha$ such that
the contribution to the expected value of
$E$ from an interval $1 \le E < c$
is always smaller than the contribution
from $c \le E< rc $,
no matter how big $c$ is.

The contribution of an interval $a\le E < c$
to the mean value of $E$ is
\begin{equation}
\int_a^c E D(E) dE \propto c^{2-\alpha} - a^{2-\alpha}.
\end{equation}
Therefore,
\begin{equation}
\int_1^c E D(E) dE \propto c^{2-\alpha} - 1,
\end{equation}
and 
\begin{equation}
\int_c^{rc} E D(E) dE \propto c^{2-\alpha} (r^{2-\alpha}-1).
\end{equation}
In order that the last integral is larger than the previous 
one it is enough that
\begin{equation}
(r^{2-\alpha}-1) c^{2-\alpha}  > c^{2-\alpha}.
\end{equation}
So, $r^{2-\alpha} > 2$ and this implies that
\begin{equation}
\label{alpha2log}
 \alpha < 2-\log_r 2.
\end{equation}
For $r=10$, the (sufficient) condition becomes $\alpha < 1.699$.
In the case of earthquake radiated energy, 
$\alpha \simeq 1+2b/3 \simeq 1.667$, 
and equation~\eqref{alpha2log} is fulfilled.
Though, slightly larger values of $\alpha$
violate the condition;
nevertheless, there is nothing special in taking $r=10$
(it is not a magical number!)
and we have that equation~\eqref{alpha2log} is fulfilled 
for a larger $r$.
For $r=2$ equation~\eqref{alpha2log} would imply $\alpha < 1$,
but this is not an acceptable exponent 
for a power-law distribution
(normalization would not be fulfilled).

\subsection*{Rigorous proof of extinction probability}

Besides graphical arguments (see Fig. \ref{iterate}), we want to provide a rigorous proof 
for the computation of the extinction probability in the Galton-Watson process, given by
\begin{equation}
P_{ext}=\lim_{t \to \infty}f^t (0),
\end{equation}
where $P_{ext}$ is properly defined only if the limit exists. To see that this is always the case, we note that
$
	Z_t =0 \implies Z_{t+1}=0
$. 
Hence, $\{Z_t=0\}\subset \{Z_{t+1}=0 \}$ and 
$
	P(Z_t =0) \leq P(Z_{t+1}=0)$, so  $f^{t}(0) \leq f^{t+1}(0)
$
or, in words, $(f^t)$ is a non-decreasing sequence. As $f([0,1]) \subset [0,1]$, we conclude that $f^t(0)$ is bounded and has a limit. To continue our proof, let us treat separately the two cases $m\leq 1$, $m>1$. Hence,

\subsubsection*{case $m \leq 1$:}
As $f(x)$ is non-convex for $x\geq 0$, it always lies above any straight line tangent to it \citep{Spivak_calculus}.
In particular, we consider the line tangent to $f(x)$ at the point $(1,1)$, and
\begin{equation} 
	f(x) > 1+m(x-1) > x.
\end{equation} 
Hence $f(x) > x$ for $0 \leq x < 1$. Also, it is straightforward to see that $f(P_{ext})=P_{ext}$,

\begin{equation}
	f\left( \lim_{t \to \infty} f^{t}(0) \right) = \lim_{t \to \infty} f(f^t(0)) = \lim_{t \to \infty}f^{t+1}(0) = \lim_{t \to \infty}f^t (0),
\end{equation}
and of course $0 \leq P_{ext} \leq 1$. So we have that $f(P_{ext})=P_{ext}$ with $0 \leq P_{ext} \leq 1$. 
Summarizing, $P_{ext}$ is a fixed point of $f(x)$ in the interval $[0,1]$, but $f(x)>x$ (strictly) in $[0,1)$. 
It is clear that the only option left is $P_{ext}=1$.

\subsubsection*{case $m > 1$:}

We will start showing that $P_{ext}\neq 1$ in this case. First, as already said, $(f^t)$ is a non-decreasing sequence. Second, as $f(x)$ is continuous and $f'(1) = m > 1$, we have that $f(x)<x$ for $x \in (1-\epsilon,1)$ for some $\epsilon > 0 $. So, $f^t(0) \notin (1-\epsilon,1)$ for all $t$ (because it would then decrease).  This means that the only way for $f^{t}(0)$ to have limit 1 is to ``jump over'' the interval $(1-\epsilon,1)$, that is, by means of some $y < 1-\epsilon$ such that $f(y)=1$. But such  $y$ cannot exist because then $f'(x) < 0$ at some point between $y$ and 1. 

Now we will see that the equation $f(x^*)=x^*$ has a unique solution in the interval $[0,1)$. There must be at least one solution because $f(0) > 0$, and $f(x)<x$ in $(1-\epsilon ,1)$ (here we are using Bolzano's theorem for $f(x)-x$). To see that this solution is unique, suppose there are two solutions, $0\leq x_1 <x_2 < 1$. As we also have $f(1)=1$, by Rolle's theorem there would exist two points $y_1, y_2$ such that $f'(y_1)=f'(y_2)=1$ and $x_1 < y_1 < x_2 < y_2 <1$, but this is impossible because $f''(x)\geq 0$ in $[0,1]$, which means that $f'(x)$ is non-decreasing and hence takes any value only once in $[0,1]$. 

So, if $P_{ext}\neq 1$ but $f(P_{ext}) =P_{ext}$, then $P_{ext}$ must be the unique solution of $f(x^* )=x^*$ in $[0,1)$. 

For the sake of rigor, we must point out that some ``pathological'' cases would need a separate treatment, such as $f(x)=x$, but those are almost never of actual interest.

\subsection*{Catalan numbers}

The Catalan numbers 
owe their name not to a Mediterranean region 
but to the French-Belgian mathematician 
from the 19th century Eug\`ene Charles Catalan.
``His'' numbers count a large variety of objects \citep{Stanley_enumerative_vol2}, 
in particular, the rooted trees that arise 
in the study of branching process when
the number of offsprings can be $0, 1$, or $2$.
We can consider a tree of size $s$
as the root (corresponding to the zero generation
of the associated branching process) plus the remaining $s-1$ nodes, 
these latter can be distributed as a varying number 
of nodes associated to the first branch, $0,1, \dots s-1$
and the rest to the second branch, $s-1, s-2, \dots 1,0$,
respectively.
Therefore, the number of trees $C_s$ of size $s$
fulfills,
\begin{equation}
C_s=C_0 C_{s-1}+C_1 C_{s-2} + \cdots + C_{s-2} C_1 + C_{s-1} C_0,
\end{equation}
where $C_0$ is taken equal to one,
as there is only one way in which a branch
can have no elements.
Note that from here we obtain 
\begin{equation}
\begin{array}{l}
C_1=(C_0)^2=1\\
C_2=2 C_0 C_1 = 2\\
C_3 = 2 C_0 C_2 +(C_1)^2=5\\
C_4 = 2 C_3 C_0 + 2 C_2 C_1 = 14\\
\end{array}
\end{equation}
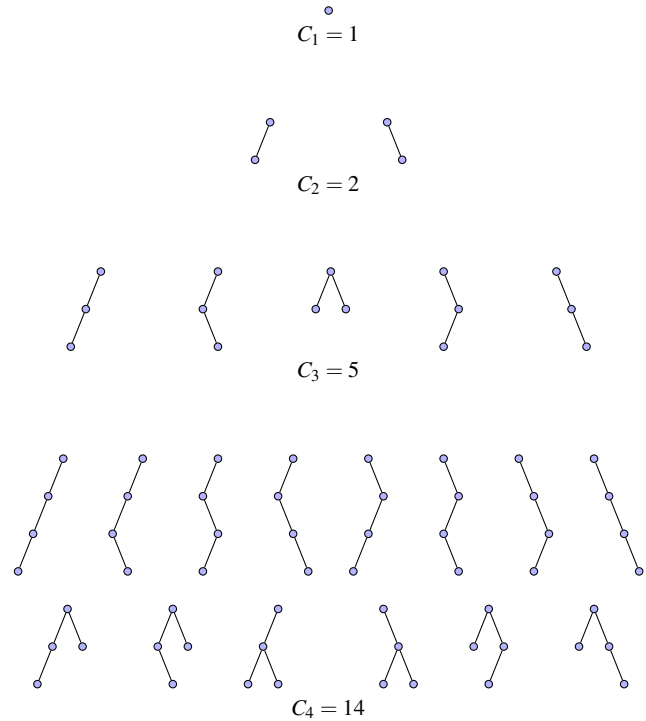
\begin{figure}[h]
\begin{center}
\begin{tikzpicture}
[
level distance=5mm,
sibling distance=4mm,
every node/.style={fill=blue!30,circle,draw,inner sep=1pt}
]
\node{}[grow=down]{
};
\end{tikzpicture}\\ $C_1=1$\\[7ex]

\begin{tikzpicture}
[
level distance=5mm,
sibling distance=4mm,
every node/.style={fill=blue!30,circle,draw,inner sep=1pt}
]
\node{}[grow=down]{
			child{node{}}
			child[missing]

};

\node[right=15mm]{}[grow=down]{
		
			child[missing]
			child{node{}}

};
\end{tikzpicture}\\ $C_2=2$\\[7ex]

\begin{tikzpicture}
[
level distance=5mm,
sibling distance=4mm,
every node/.style={fill=blue!30,circle,draw,inner sep=1pt}
]
\node{}[grow=down]{
		child{node{}
			child{node{}}
			child[missing]
		}
		child[missing]
	
};

\node[right=15mm]{}[grow=down]{
		child{node{}
			child[missing]
			child{node{}}
		}
		child[missing]

};
\node[right=30mm]{}[grow=down]{

		child{node{}}
		child{node{}}
	
};
\node[right=45mm]{}[grow=down]{
		child[missing]
		child{node{}
			child{node{}}
			child[missing]
		}

};

\node[right=60mm]{}[grow=down]{
		child[missing]
		child{node{}
			child[missing]
			child{node{}}
		}

};
\end{tikzpicture}	\\$C_3=5$
 \\[7ex]

\begin{tikzpicture}
[
level distance=5mm,
sibling distance=4mm,
every node/.style={fill=blue!30,circle,draw,inner sep=1pt}
]
\node{}[grow=down]{
	child{node{}
		child{node{}
			child{node{}}
			child[missing]
		}
		child[missing]
	}
	child[missing]
};

\node[right=10mm]{}[grow=down]{
	child{node{}
		child{node{}
			child[missing]
			child{node{}}
		}
		child[missing]
	}
	child[missing]
};

\node[right=20mm]{}[grow=down]{
	child{node{}
		child[missing]
		child{node{}
			child{node{}}
			child[missing]
		}
	}
	child[missing]	
};

\node[right=30mm]{}[grow=down]{
	child{node{}
		child[missing]
		child{node{}
			child[missing]
			child{node{}}
		}
	}
	child[missing]
};

\node[right=40mm]{}[grow=down]{
	child[missing]
	child{node{}
		child{node{}
			child{node{}}
			child[missing]
		}
		child[missing]
	}
};

\node[right=50mm]{}[grow=down]{
	child[missing]
	child{node{}
		child{node{}
			child[missing]
			child{node{}}
		}
		child[missing]
	}
};

\node[right=60mm]{}[grow=down]{
	child[missing]
	child{node{}
		child[missing]
		child{node{}
			child{node{}}
			child[missing]
		}
	}
};

\node[right=70mm]{}[grow=down]{
	child[missing]
	child{node{}
		child[missing]
		child{node{}
			child[missing]
			child{node{}}
		}
	}
};

\node[below=20mm,right=0mm]{}[grow=down]{
	child{node{}
		child{node{}}
		child[missing]
	}
	child{node{}}
};

\node[below=20mm,right=14mm]{}[grow=down]{
	child{node{}
		child[missing]
		child{node{}}
	}
	child{node{}}
};

\node[below=20mm,right=28mm]{}[grow=down]{
	child{node{}
		child{node{}}
		child{node{}}
	}
child[missing]
};

\node[below=20mm,right=42mm]{}[grow=down]{
	child[missing]
	child{node{}
		child{node{}}
		child{node{}}
	}

};
\node[below=20mm,right=56mm]{}[grow=down]{
	child{node{}}
	child{node{}
		child{node{}}
		child[missing]
	}
};

\node[below=20mm,right=70mm]{}[grow=down]{
	child{node{}}
	child{node{}
		child[missing]
		child{node{}}
	}
};
\end{tikzpicture}\\ $C_4=14$
\end{center}
\caption{The number of rooted trees with no more than two branches per node is shown, up to size $s=4$.
The number of such trees of a given size is given by $C_s$, the $s$-th Catalan number.}
\label{trees}
\end{figure}
and so on this simple formula generates all Catalan numbers. The curious reader can check Figure \ref{trees}, where all possible rooted trees with no more than two branches per node, of size up to 4, are shown.

If we want a closed expression for these numbers, 
we may define a generating function
\begin{equation}
h(x)=C_0+C_1 x +C_2 x^2 +\dots =\sum_{s=0}^\infty C_s x^s.
\end{equation}
One can obtain an expression for $h(x)$ just
using the properties of the Catalan numbers \citep{Wilf}.
First, let us calculate 
\begin{align*}
[h(x)]^2 &= \left[\sum_{s=0}^{\infty}C_s x^s \right]^2=
\sum_{i,j=0}^{\infty}C_{i}C_{j}x^{i+j}= \\&=
\sum_{s=0}^{\infty}\underbrace{\left[ \sum_{i+j=s}C_i C_j \right]}_{C_{s+1}} x^s
= \frac 1 x {\sum_{s=0}^{\infty}C_{s+1}x^{s+1}} = \frac  {h(x)-C_0}{x}
\end{align*}
As we know that $C_0=1$, we end up with a quadratic equation for $h(x)$, namely,
\begin{equation} 
x [h(x)]^2 - h(x) +1 =0,
\end{equation} 
which allows us to isolate $h(x)$,
\begin{equation} 
h(x)=\frac{1\pm \sqrt{1-4x}}{2x}.
\end{equation} 
One of both functions (depending on the $\pm$ sign)
is then the generating function of the Catalan numbers.
We are going to recover these numbers from its generating function.
First, one needs the Taylor expansion of $\sqrt{1-x}$ around $x=0$,
which is
\begin{equation}
\sqrt{1-x}=1 -\frac x 2  -\frac 1 4 \frac {x^2}{2!} -\frac 3 8 \frac{x^3}{3!}-\dots
=1 -\frac x 2 -\sum_{s=1}^\infty \frac{(2s-1)!!}{2^{s+1}(s+1)!} x^{s+1},
\end{equation}
where, remember, $n!!=n (n-2)\cdots 1$, and so,
\begin{equation}
\sqrt{1-4x}=1 -2 x -\sum_{s=1}^\infty \frac{(2s-1)!! 2^{s+1}}{(s+1)!} x^{s+1}.
\end{equation}
Then, substituting in $h(x)$, one can realize that only the minus sign can
correspond to a generating function, and
\begin{equation}
h(x)= 1 + \frac 1 {2x} 
\sum_{s=1}^\infty \frac{(2s-1)!! 2^{s+1}}{(s+1)!} x^{s+1}=
1+
\sum_{s=1}^\infty \frac{(2s-1)!! 2^{s}}{(s+1)!} x^{s},
\end{equation}
from where we obtain a first expression for the Catalan numbers,
\begin{equation}
C_s= \frac{(2s-1)!! 2^{s}}{(s+1)!}  \, \mbox { for } s \ge 1.
\end{equation}
A more comfortable formula can be obtained using that
\begin{equation}
(2s)! = (2s)!! (2s-1)!! = s! 2^s (2s-1)!!,
\end{equation} 
and then one finds,
\begin{equation}
C_s=\frac{(2s)!}{s! (s+1)!} = \frac 1 {s+1} 
\left(
\begin{array}{c}
2s \\
s \\
\end{array}
\right),
\end{equation}
the standard expression for the Catalan numbers,
now valid for all $s\ge 0$.

\subsection*{Normalization and non-normalization of the total size distribution}

We are going to illustrate how the total size probability distribution, $P(S=s)$,
is only normalized in the subcritical and critical cases.
We use the binomial distribution for the distribution of the
number of offsprings, with $k=0,1$ and $2$.
From the main text, we know that
\begin{equation}
P(S=s) = \frac 1 {s+1} 
\left(
\begin{array}{c}
2s \\ s \\
\end{array}
\right)
p^{s-1}(1-p)^{s +1}
\, \mbox{ with } \, 
s=1,2,\dots
\end{equation}
It can be checked, using the generating function
of the Catalan numbers, that 
this expression is normalized for $p\le 1/2$ but not for $p> 1/2$.
In order to see this,
let us first consider the generating function of the Catalan numbers, 
derived in the previous subsection of the Appendix,
\begin{equation}
h(x)=\sum_{s=0}^\infty C_s x^s=\frac{1-\sqrt{1-4x}}{2x}.
\end{equation}
Then, introducing $q=1-p$,
\begin{equation}
\sum_{s=1}^\infty P(S=s)=
\frac{q}{p}\sum_{s=1}^\infty C_s (pq)^s=
\frac{q}{p} \left(h(pq)-1\right),
\end{equation}
and using the expression for $h(x)$,
\begin{equation}
h(pq)=\frac{1-\sqrt{1-4pq}}{2pq}
=\frac{1-\sqrt{(1-2p)^2}}{2pq}
=\frac{1-|1-2p|}{2pq}.
\end{equation}
We can distinguish two cases, first, $p\le 1/2$, for which,
\begin{equation}
h(pq)-1=\frac 1 q -1 = \frac p q =\frac{\min(p,q)}{\max(p,q)},
\end{equation}
and for the opposite case, $p\ge 1/2$,
\begin{equation}
h(pq)-1=\frac 1 p -1 = \frac q p  =\frac{\min(p,q)}{\max(p,q)}.
\end{equation}
Therefore,
\begin{equation}
\sum_{s=1}^\infty P(S=s)=\frac q p
\frac{\min(p,q)}{\max(p,q)}=
\left\{
\begin{array}{cl}
1 & \mbox{ for } p\le 1/2 \\
\left(\frac q p\right)^2
& \mbox{ for } p\ge 1/2 \\
\end{array}
\right.
\end{equation}
Remembering the results for the extinction probability for
the binomial distribution,
\begin{equation}
\sum_{s=1}^\infty P(S=s)=P_{ext},
\end{equation}
which obviously is not normalized for $p > 1/2$.
We could also have arrived to the same result using, 
not the generating function of the Catalan numbers, 
but the generating function $g(x)$ of the size $S$.

\subsection*{Stirling's Approximation}

Usually, Stirling's formula is demonstrated
by means of the Euler-Maclaurin formula.
However, if one knows some elementary properties of the
gamma distribution, Stirling's formula arises
almost spontaneously, by means of
a probabilistic trick.

 Remember that the factorial is associated to the gamma function,
$n! = \Gamma(n+1)$, which is defined as
\begin{equation}
\Gamma(\gamma)=\int_0^\infty y^{\gamma-1} e^{-y} dy
\end{equation}
for $\gamma > 0$ \citep{Abramowitz}.
This allows to introduce the gamma distribution \citep{Durrett},
with probability density given by
\begin{equation}
\frac 1 {\Gamma(\gamma)} y^{\gamma-1} e^{-y}
\end{equation}
for $y \ge 0$ (and zero otherwise),
and with mean $\gamma$ and variance $\gamma$.

\begin{figure}
\includegraphics[width=\textwidth]{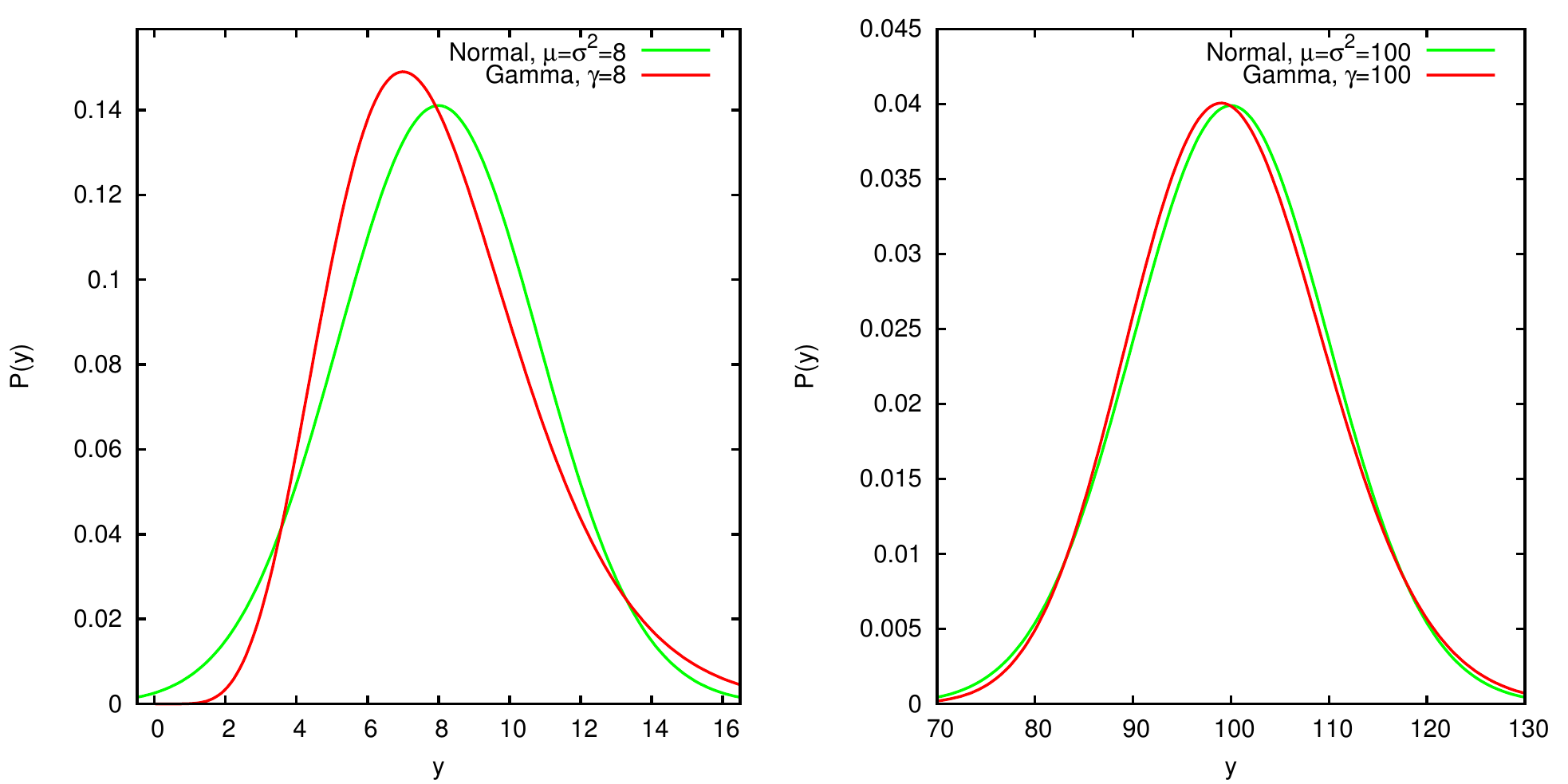}
\caption{Approaching of the normal distribution by the gamma distribution, 
adding 8 and 100 exponentials, respectively.
The central limit theorem allows the derivation of Stirling's approximation.}
\label{gamma}
\end{figure}

It turns out that the gamma distribution
arises as a sum of a number $\gamma$ of independent exponential
random variables, each with density $e^{-y}$
(this can be easily demonstrated through successive convolutions
of the exponentials, see \cite{Durrett}).
But using the central limit theorem, the gamma distribution will converge,
in the limit $\gamma\rightarrow \infty$,
to a normal distribution (see Fig. \ref{gamma}), with mean $\mu$ 
and standard deviation $\sigma$
(in this case the notation is different to the rest of the chapter).

Then, it will be possible to 
transform the gamma function
into a Gaussian integral. 
Indeed,
\begin{equation}
n!=\Gamma(n+1)=\int_0^\infty y^{n} e^{-y} dy \rightarrow 
C \int_0^\infty \exp\left(-\frac{(y-\mu)^2}{2 \sigma^2}\right) dy.
\end{equation}
The key point is to find the value of $C$ for which both functions
overlap. This happens around the mean or the mode of both distributions,
corresponding, respectively, to $y=\gamma=n+1\simeq n$ and $y=\mu$.
Substituting both values in
\begin{equation}
y^n e^{-y} = C \exp\left(-\frac{(y-\mu)^2}{2 \sigma^2}\right) 
\end{equation}
we get 
\begin{equation}
C=\left(\frac n e\right)^n 
\end{equation}
and therefore, looking for the normal probability density
inside the integral,
\begin{equation}
n!=\Gamma(n+1)\rightarrow
\sqrt{2 \pi} \sigma C  
\int_0^\infty \frac 1 { \sqrt{2 \pi} \sigma}
\exp\left(-\frac{(y-\mu)^2}{2 \sigma^2}\right) dy.
\end{equation}
The value of $\sigma$ is obtained from $\sigma^2 =\gamma=n+1$
(for independent random variables the variance of a sum
is the sum of variances, which is one for each exponential distribution in our sum).
Substituting, and replacing the lower integration limit by $-\infty$,
due to the fact that the standard deviation $\sigma\simeq \sqrt{n}$ is much smaller than the mean $\mu\simeq n$,
one obtains,
\begin{equation}
n! \sim \sqrt{2 \pi n} \left(\frac n e\right)^n,
\end{equation}
valid, remember, in the limit $n \rightarrow \infty$.
This proof has some parts in common with the more elaborated one of \cite{Khan_Stirling}
and less resemblance with that of \cite{van_den_Berg}.

\begin{acknowledgement}
We would like to dedicate this work to the colorful scientist
Per Bak, in the 25 years of his invention of self-organized criticality
and in the 10th anniversary of his untimely death.
The chapter originates, in part, from a lecture 
that one of the authors gave at the 2011 Fall Meeting of the
{\it American Geophysical Union}.
In this regard, we thank Armin Bunde,
and also
Tom Davis, for making his notes on the Catalan numbers publicly available
on the Internet, and 
Anna Deluca and Gunnar Pruessner, for discussions.
Cec\'{\i}lia M. Clos provided valuable graphical-design assistance.
Funding has come from Spanish projects
FIS2009-09508
and 2009-SGR-164.

\end{acknowledgement}

\bibliographystyle{apalike}
\bibliography{biblio}

\end{document}